\documentclass[a4paper,fleqn]{cas-dc}

\usepackage[authoryear]{natbib}
\usepackage[utf8]{inputenc}
\usepackage{graphicx}
\usepackage{booktabs}
\usepackage{makecell}
\usepackage{array}
\usepackage{multirow}
\usepackage{tabularx}
\usepackage{colortbl}
\usepackage{adjustbox}
\usepackage{rotating}
\usepackage{placeins}
\usepackage{indentfirst}
\usepackage{ragged2e}
\usepackage{balance}
\usepackage{enumitem}
\usepackage{amsmath,amssymb}
\usepackage{pifont}
\usepackage{fontawesome}
\usepackage{newunicodechar}
\newunicodechar{）}{)}
\usepackage{xcolor}
\definecolor{cyan}{HTML}{009aa8}
\usepackage[most]{tcolorbox}
\usepackage{framed}
\definecolor{Gray}{gray}{0.9}
\definecolor{shadecolor}{gray}{0.95}
\usepackage{tikz}
\usetikzlibrary{trees,positioning,shapes,shadows,arrows}
\usepackage{hyperref}
\usepackage{xurl}

\tikzset{
  basic/.style  = {draw, text width=2cm, drop shadow, font=\sffamily, rectangle},
  root/.style   = {basic, rounded corners=2pt, thin, align=center, fill=white},
  level-2/.style = {basic, rounded corners=6pt, thin,align=center, fill=white, text width=3cm},
  level-3/.style = {basic, thin, align=center, fill=white, text width=1.8cm}
}

\newcommand{\piechartfont}{\normalfont\fontfamily{phv}\fontseries{m}\fontshape{n}\selectfont}
\newcommand{\pieslicelabelfont}{\piechartfont\fontsize{5.6pt}{5.8pt}\selectfont}
\newcommand{\pielabeltext}[2]{{\normalfont\fontfamily{phv}\fontseries{m}\fontshape{n}\fontsize{5.6pt}{5.8pt}\selectfont #1, #2\%}}
\newcommand{\tikzlegendfont}{\piechartfont\scriptsize}

\newcommand{\surveyuser}[1]{\mbox{\faUser~#1}}
\newif\ifcleanversion
\cleanversiontrue
\ifcleanversion
  \newcommand{\rev}[1]{#1}
  
  \newcommand{\todo}[1]{}
  \newcommand{\revcolor}{}
\else
  \newcommand{\rev}[1]{\textcolor{blue}{#1}}
  
  \newcommand{\todo}[1]{{\color{red}\textbf{[TODO: #1]}}}
  \newcommand{\revcolor}{\color{blue}}
\fi

\begin{document}

\let\WriteBookmarks\relax
\def\floatpagepagefraction{1}
\def\textpagefraction{.001}

\shorttitle{Using LLMs in Software Design}
\shortauthors{Y Wang et al.}

\title [mode = title]{Using LLMs in Software Design: An Empirical Study of GitHub and A Practitioner Survey}

\author[1]{Yifei Wang}
\ead{whiten@whu.edu.cn}
\credit{Conceptualization, Investigation, Data curation, Formal analysis, Writing - Original draft preparation}

\author[1]{Ruiyin Li}
\cormark[1]
\ead{ryli_cs@whu.edu.cn}
\credit{Methodology, Validation, Data curation, Writing - review and editing}

\author[1]{Peng Liang}
\cormark[1]
\ead{liangp@whu.edu.cn}
\credit{Conceptualization, Methodology, Data curation, Supervision, Writing - review and editing}

\author[1]{Yangxiao Cai}
\ead{yangxiaocai@whu.edu.cn}
\credit{Investigation, Data curation, Writing - review and editing}
\address[1]{School of Computer Science, Wuhan University, China}

\author[2]{Zengyang Li}
\ead{zengyangli@ccnu.edu.cn}
\credit{Methodology, Writing - review and editing}
\address[2]{School of Computer Science \& Hubei Provincial Key Laboratory of Artificial Intelligence and Smart Learning, Central China Normal University, China}

\author[3]{Mojtaba Shahin}
\ead{mojtaba.shahin@rmit.edu.au}
\credit{Methodology, Writing - review and editing}
\address[3]{School of Computing Technologies, RMIT University, Australia}

\author[4]{Arif Ali Khan}
\ead{arif.khan@oulu.fi}
\credit{Methodology, Writing - review and editing}
\address[4]{M3S Research Unit, University of Oulu, Finland}

\author[5]{Qiong Feng}
\ead{qiongfeng@njust.edu.cn}
\credit{Methodology, Writing - review and editing}
\address[5]{School of Computer Science, Nanjing University of Science and Technology, China}

\cortext[cor1]{Corresponding author.}

\begin{abstract}
Recent advancements in Large Language Models (LLMs) have demonstrated significant potential across software engineering tasks, including software design, an area traditionally regarded as highly dependent on human expertise and judgment. However, limited research has examined how LLMs are used in software design, as well as the associated benefits and drawbacks. This paper \rev{addresses} this gap by empirically investigating how software developers \rev{use LLMs in software design}. We \rev{conducted} a mixed-methods study, combining a mining study of 291 developer-ChatGPT conversations shared on GitHub with a survey of 65 software practitioners. \rev{From the mined conversations, we identified} nine categories of design tasks supported by ChatGPT, including architecture design, data model design, and the use of design patterns. We further characterize developer-ChatGPT interactions, showing that developers \rev{in the mined conversations} primarily use ChatGPT for knowledge acquisition and design-related code generation, with most tasks situated at the detailed design level. The \rev{survey participants reported} seven key benefits, such as better technology selection and early detection of design flaws, \rev{and six limitations, including lengthy outputs, inexecutable or incorrect code, and dependence on project context}. These findings provide an evidence-based characterization of current LLM use in software design from both open-source and practitioner perspectives. 
\end{abstract}


\maketitle

\begin{sloppypar}

\section{Introduction}\label{sec:Introduction}
Recent advances in Large Language Models (LLMs) have rapidly expanded both practice and tooling in Software Engineering (SE). Several secondary studies summarize that LLMs have been explored across a broad range of SE activities, including requirements, design, coding, testing, maintenance, and documentation~\citep{hou2024llm4se_slr,Fan2023LLMSurvey}. Meanwhile, LLM-powered development tools have emerged to integrate LLMs throughout the development workflow. For example, \citet{qian-etal-2024-chatdev} proposed a multi-agent framework, ChatDev, that coordinates role-specific LLMs to simulate a virtual software development organization across multiple lifecycle phases. However, existing empirical studies, tools, and benchmarks largely concentrate on code-level tasks (e.g., code generation, completion, and bug fixing)~\citep{swebench, hou2024llm4se_slr}. By contrast, higher-level SE activities that require reasoning over abstractions and trade-offs, particularly requirements engineering and software design, are comparatively less studied~\citep{Fan2023LLMSurvey,hou2024llm4se_slr}. In particular, \citet{hou2024llm4se_slr} report that \rev{studies explicitly focused on software design accounted for only 0.92\% of the primary studies they classified by SE activity between January 2017 and January 2024}. This finding suggests that LLM support for software design remains underexplored, potentially because design outcomes are more difficult to specify and evaluate than code-level outputs and often depend on rich, project-specific context.

This lack of empirical understanding of LLMs for software architecture and design is particularly critical, as design decisions shape a system's high-level structure and long-term quality attributes~\citep{garlan1994intro}. Architectural decisions are also costly to revise once a system grows; during evolution, inconsistencies between the intended design and implemented structures can accumulate, resulting in architectural erosion and increased maintenance effort~\citep{li2022understanding}. However, existing studies primarily focus on capability demonstrations or small-scale examples, providing limited empirical evidence on how developers actually use LLMs for architecture and design in real-world settings~\citep{schmid2025slr}. As a result, many questions remain unclear; for example, what design problems developers actually attempt to solve with LLMs in practice; how they iteratively frame and refine design inquiries; and where perceived value and friction arise when LLMs are used to support software design.

Addressing these questions requires evidence from practical settings. Existing practice-oriented work has begun to examine the adoption of LLMs in SE; for example, \citet{StateOfPractice} surveyed 15 European software companies to understand their use of LLMs, perceived benefits, challenges, and future adoption plans in SE and software architecture. However, such studies provide limited evidence on how developers interact with ChatGPT when addressing concrete software design tasks. Evidence grounded in developers' multi-turn interactions with LLMs and practitioners' opinions and experiences can reveal what design tasks are actually attempted with LLMs, how developers structure prompts and iterations, and where perceived value and friction arise. Such understanding can complement prior studies~\citep{StateOfPractice, schmid2025slr} by linking observed interaction behaviors with practitioners' perceived benefits and limitations.

We aim to empirically investigate how developers utilize LLMs in practical software design, what benefits they perceive, and what limitations they encounter. To this end, we conducted a mixed-methods study, combining a mining study and a survey study, to characterize developer-LLM interactions around software design issues and to examine the benefits and limitations perceived by developers when using LLMs for such issues. Building upon previous research on shared ChatGPT conversation records in GitHub~\citep{Usage_of_ChatGPT_in_Github}, we constructed a dataset~\citep{replpack} comprising 291 records of shared conversations between developers and ChatGPT related to software design in GitHub, spanning from May 2023 to January 2025. We labeled the conversations in the dataset and adopted the Open Coding and Constant Comparison method~\citep{CC2014} to derive emergent categories from the qualitative data for describing design tasks that can be addressed by ChatGPT and the ways developers interact with ChatGPT to address such tasks. Furthermore, based on the results of the mining study, we formulated a questionnaire and conducted a survey study with 65 software \rev{practitioners} who utilize LLMs for software design tasks.

\rev{We use the following terminology consistently throughout the paper. We use \textit{developers} to refer to the people whose shared conversations we mined, because the activity we observe from them is development-related activity on GitHub. We use \textit{participants} to refer to the 65 survey respondents, and \textit{practitioners} when a statement applies to both groups or to the broader population to which they belong. This distinction is particularly important for the survey, whose respondents occupy a range of roles beyond development, including architects, testers, requirements engineers, consultants, team leaders, and project managers (see Section~\ref{sec:Data Conducting surve}).}

Our research has unveiled the following \textbf{key findings}: \textbf{(1)} ChatGPT has been employed in a variety of tasks related to software design, such as architecture design and data model design. \textbf{(2)} ChatGPT is primarily used as a knowledge resource, typically involving multi-turn conversations \rev{with a median of five rounds}. \textbf{(3)} The most valued benefits of using LLMs in software design are their efficiency as a search-engine alternative and their capability for the early detection of design flaws. \textbf{(4)} The most significant challenge reported by \rev{the survey participants} is that LLMs produce excessively lengthy outputs.

The rest of this paper is organized as follows: Section~\ref{sec:RelatedWork} presents the related work. Section~\ref{sec:Study Design} describes the research design of this study. Section~\ref{sec:Results} provides the results of the study, which are further discussed in Section~\ref{sec:Discussions & Implications}. The potential threats to the validity of the study are clarified in Section~\ref{sec:Threats}, and Section~\ref{sec:Conclusions} concludes this work with future directions.

\section{Related Work}\label{sec:RelatedWork}
In this section, we first review the studies on LLMs for software engineering tasks (Section~\ref{sec:LLMsForSE}) and then how LLMs are used for software design (Section~\ref{sec:LLMsForSDIssues}). We finally position our study in relation to prior work (Section~\ref{sec:Conclusive Summary}).

\subsection{LLMs for Software Engineering}\label{sec:LLMsForSE}

Within Software Engineering (SE), LLMs are increasingly adopted across the software development lifecycle (SDLC), including requirements engineering, design, implementation, testing, maintenance, and project management. A growing body of secondary studies (e.g.,~\citep{hou2024llm4se_slr,schmid2025slr}) indicates that the literature has expanded rapidly but remains unevenly distributed across SE tasks: implementation-centric work (e.g., code generation, code completion, and bug fixing) dominates, whereas evidence for higher-level activities (e.g., architecture and design) is comparatively limited and often relies on smaller-scale evaluations or case studies. Meanwhile, the community has begun to operationalize evaluation through realistic benchmarks and task suites that employ end-to-end development scenarios, such as resolving real GitHub issues within their repository context~\citep{swebench}.

The dominant body of evidence on LLMs for SE is concentrated on developer-assistance tools for coding. For example, controlled experiments and large-scale observational studies on GitHub Copilot~\citep{peng2023copilot,githubcopilot_research} report improvements in task completion speed and developer-perceived productivity, while highlighting caveats such as the need for careful verification of content generated by \rev{Copilot}, variability of \rev{Copilot's} capabilities across tasks and experience levels, and the risk of overtrusting \rev{LLM-based} tools. Beyond interactive code completion, research has explored the use of LLMs for more complex forms of automation, including issue resolution and agentic workflows that integrate tool execution, environment feedback, and iterative planning (e.g.,~\citep{swebench,react,toolformer,agentless}). These LLM-based systems show promise in reducing execution friction; however, they also reveal practical limitations (e.g., hallucinated actions, brittle plans, and evaluation or replication challenges), which complicate their deployment in safety- and correctness critical SE contexts.

Overall, existing work demonstrates that LLMs can meaningfully support a range of SE tasks; nevertheless, the strongest evidence remains concentrated on code-level assistance (e.g., code generation). This motivates us to conduct an empirical investigation of how developers actually use LLMs for software design in practice, which specific design tasks have been supported by LLMs, and what benefits and limitations are identified when applying LLMs to software design.

\subsection{LLMs for Software Design}\label{sec:LLMsForSDIssues}
Compared with code-level assistance, using LLMs for software design and architecture requires models to operate over higher-level abstractions (e.g., components, interfaces, and constraints) and to connect these abstractions to project-specific contexts. Recent evidence suggests that research in this area is growing but remains fragmented: a systematic literature review focusing on software architecture reports that LLMs have been explored for tasks such as generating architecture design from requirements, extracting architectural decisions, and detecting architecture patterns; yet evaluations are often small-scale, and the degree of automation varies widely. In particular, the review highlights that important architecture tasks such as evaluating quality attributes (e.g., evolvability) and architecture conformance checking are largely absent, and that many studies lack baseline comparisons, indicating the need for more rigorous benchmarking and evaluation~\citep{schmid2025slr}.

A prominent stream of research investigates requirement-to-design translation, in which LLMs convert natural-language requirements into structured design artifacts. Recent studies evaluate LLMs (and LLM-based agents) on generating UML diagrams from textual requirements or user stories, proposing prompting or agent frameworks to analyze structural and semantic deviations in the design~\citep{acm_uml_eval_2024,nomad_uml_2025,cot_classdiagram_2024}. These results collectively indicate that LLMs can produce usable initial design, but correctness and completeness are sensitive to the specificity of prompts, domain assumptions, and the model's ability to preserve constraints across diagram entities and relationships~\citep{acm_uml_eval_2024,nomad_uml_2025}.

Another active direction is architectural and design knowledge extraction and analysis, including reusable architectural knowledge (e.g., patterns) as well as project-specific design decisions and rationale, often captured as Architecture Decision Records (ADRs). For example, an empirical study on design pattern recognition shows that the performance of LLM-based design pattern recognition varies by pattern type and depends on capturing semantic and contextual information in addition to syntactic or structural properties of the code (e.g., implementation context, keywords, and implementation style)~\citep{dprLLMStudy}. Complementary work such as ArchMind combines architectural knowledge about patterns with information about the system under design to support architectural decision making and ADRs generation; it also reports that generated ADRs may miss system-specific details when the system context is insufficient~\citep{diazpace2024archmind}. \citet{zhou2025using} used LLMs to generate and recover design rationale for architecture decisions. These observations suggest that effective LLM-based design assistance depends strongly on how design constraints and project context are supplied to the model. For instance, prompt-pattern catalogs treat prompts as reusable artifacts that encode constraints and guidance for tasks such as requirements analysis and system design~\citep{white2023promptpatterns}. In addition, recent reviews of generative AI for software architecture report that some approaches retrieve project or domain artifacts (e.g., architectural documentation or ADR repositories) and incorporate them into model inputs to provide additional context for generation~\citep{esposito2026genai_sa}. Recent empirical studies have also examined LLMs for generating architecture-specific artifacts. \citet{Dhar2024ADD} explored whether LLMs can generate architectural design decisions from decision contexts and found that state-of-the-art models can generate relevant decisions but still fall short of human-level performance. \citet{Arun2025ArchitecturalComponents} investigated LLM-based generation of serverless architectural components and the results showed that generated components depend on the provided system context and require evaluation using correctness and quality metrics.

Despite encouraging results, using LLMs for design inherits the core limitations of LLMs, such as hallucination, non-determinism, and context constraints, and introduces additional risks, as design decisions have long-term consequences. Recent evidence on AI-assisted programming also cautions that productivity gains are not guaranteed for experienced developers on complex repositories: a randomized controlled trial reports that allowing early-2025 AI tools slowed down experienced open-source developers on their own tasks, partly due to verification and integration overhead~\citep{metr2025rct}. These findings motivate empirical studies that examine how developers actually interact with LLMs in design contexts, what they ask, how they iterate, and what benefits and limitations arise in practice.

\subsection{Conclusive Summary}\label{sec:Conclusive Summary}
Prior work shows that LLMs can support a broad spectrum of SE activities, yet the strongest evidence remains concentrated on implementation-centric assistance and issue-resolution scenarios. Research on software design and architecture is growing, with empirical studies examining tasks such as architecture design generation from requirements, generation of architectural design decisions and ADRs, detection and classification of architectural decisions or tactics in code, extraction of design rationale, and generation of architectural components~\citep{schmid2025slr,Dhar2024ADD,Arun2025ArchitecturalComponents,zhou2025using}. However, much of this evidence still comes from capability-oriented evaluations, sparse datasets, or narrow task scenarios. \rev{Moreover, design decisions shape a system's long-term structure and quality attributes~\citep{garlan1994intro,li2022understanding}, and recent evidence from AI-assisted programming indicates that verification and integration overhead can offset productivity gains even for experienced developers~\citep{metr2025rct}.} Therefore, developers should use LLMs for software design in ways that align with project-specific constraints and remain aware of the boundaries of their capabilities and limitations.

These observations motivate our study to focus on \textit{how developers actually use LLMs for software design in the wild}. By analyzing publicly shared developer-ChatGPT conversations related to software design and triangulating with a complementary practitioner survey, our work complements previous capability-oriented studies by providing empirical evidence on real-world usage characteristics of developer-ChatGPT interaction, as well as practitioners' perceived benefits and limitations of utilizing LLMs in software design.

\section{Study Design}\label{sec:Study Design} 
\begin{figure*}[htbp]
\centering
\includegraphics[width=1\textwidth]{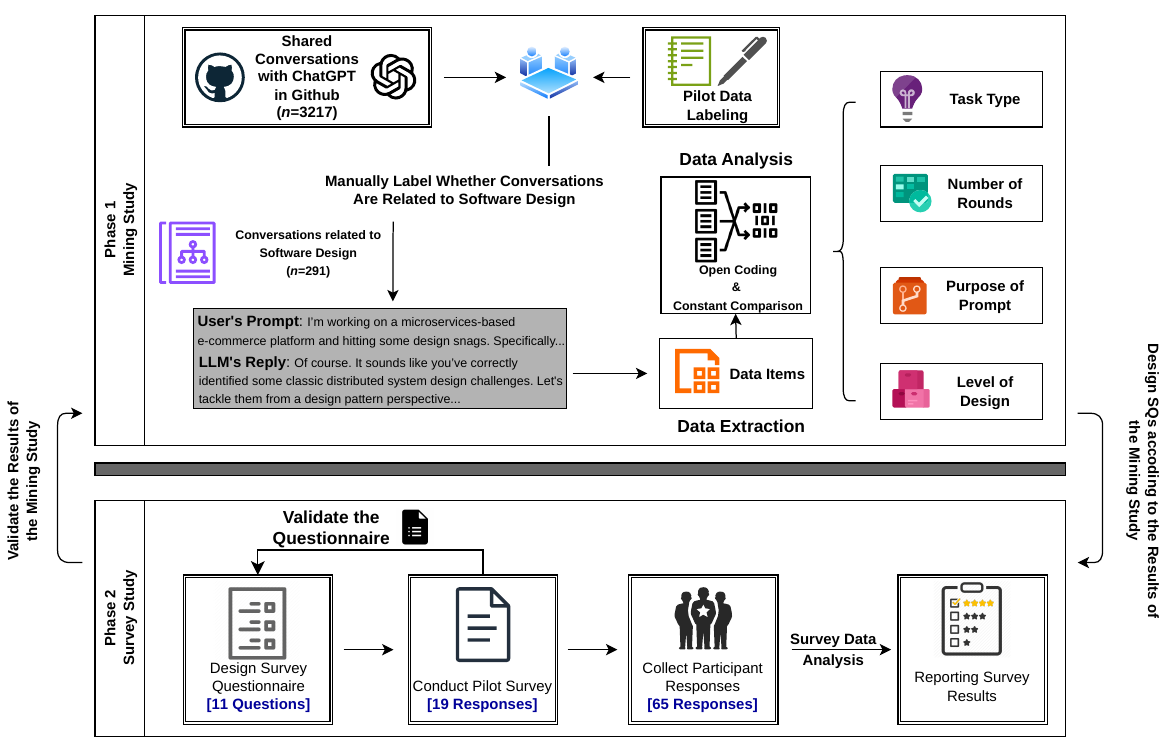}
\caption{Overview of the mixed-methods research process}
\label{fig:OverviewOfProcess}
\end{figure*}

We employed a mixed-methods approach, combining a mining study and a survey study, to explore how developers use ChatGPT and LLMs to address software design tasks in practice. Specifically, we adopted the Open Coding and Constant Comparison method~\citep{CC2014} for the conversations between developers and ChatGPT related to software design. Then, based on the derived categories of design tasks supported by ChatGPT and developers' usage of ChatGPT in software design from the mining study, we conducted a survey with 65 valid responses to \rev{complement} the derived categories and further explore the benefits and limitations of using LLMs in software design. In this section, we first present our four Research Questions (RQs), and then we detail the research process (see Figure~\ref{fig:OverviewOfProcess}) and the methods utilized for data collection, labeling, extraction, and analysis.

{\revcolor
Figure~\ref{fig:OverviewOfProcess} summarizes the research process. The mining study comprises four stages: data collection from five GitHub sources, labeling against the inclusion and exclusion criteria, extraction of data items D1$\sim$D4, and open coding and constant comparison to derive the task categories for answering RQ1 and the prompt-purpose and design-level categories for answering RQ2 (Sections~\ref{sec:Data Collection} and~\ref{sec:Data Extraction and Analysis}). These stages produce 3,217 collected conversations, 291 design-related conversations, and 376 design task instances.

The survey consists of three stages: questionnaire development, piloting and distribution, and data analysis. The categories identified in the mining study were used as candidate options for survey questions SQ4, SQ6, and SQ7, while SQ8$\sim$SQ10 elicited participants' reported benefits and limitations. The questionnaire was distributed to 1,130 potential participants recruited through GitHub and LinkedIn, yielding 65 valid responses. Closed-ended and semi-open-ended responses were analyzed descriptively, whereas free-text responses were analyzed using open coding and constant comparison (Sections~\ref{sec:Questionnaire Creating and Participants recruitment}--\ref{sec:Data Conducting surve}).

The questionnaire serves as a bridge between the two studies. Survey questions SQ4, SQ6, and SQ7 complement the mining study findings for RQ1 and RQ2, whereas RQ3 and RQ4 are addressed exclusively through the survey. Because the response options were derived from the mining study, the survey did not independently validate the resulting taxonomy.
}

\subsection{Research Questions}\label{sec:RQs}
Our study aims to understand how practitioners utilize LLMs in software design and identify the associated benefits and challenges to guide developers using LLMs more effectively. To achieve this goal, we formulate four Research Questions (RQs). RQ1 and RQ2 are answered through an exploratory mining study by analyzing the conversations between developers and ChatGPT related to software design and are further \rev{complemented by} a survey study that included 65 responses. The results of RQ1 and RQ2 can help identify the specific software design tasks supported by ChatGPT and the usage behavior of developers. RQ3 and RQ4 are answered through the survey study. The results of RQ3 and RQ4 can help us understand the benefits and limitations of using LLMs in software design tasks.

\begin{tcolorbox}
\textbf{RQ1: What specific software design tasks have been supported by ChatGPT?}
\end{tcolorbox}
\textbf{Rationale}: This RQ aims to understand the extent to which ChatGPT is being used in various tasks related to software design, such as architecture design, component design, and interface design. The answer \rev{to} this RQ can provide a landscape of the practical applications of ChatGPT in software design and inform future research directions. 

\begin{tcolorbox}
\textbf{RQ2: How \rev{is ChatGPT} used in software design?}
\end{tcolorbox}
\textbf{Rationale}: This RQ intends to explore the specific ways in which developers employ ChatGPT in their design processes. By examining the methods and approaches used, we can gain insights into the practical applications of ChatGPT, such as generating design patterns, suggesting architectural improvements, or assisting in the creation of design documents. Understanding these usage characteristics will provide a clearer picture of how developers interact with ChatGPT to enhance the software design process. 

\begin{tcolorbox}
\textbf{RQ3: What are the benefits of using LLMs in software design?}
\end{tcolorbox}
\textbf{Rationale}: This RQ seeks to identify and characterize the specific benefits that developers perceive from using LLMs in software design. These benefits could include improved design quality, increased development efficiency, reduced development time, or enhanced collaboration among team members. By identifying these benefits, we can gain a deeper understanding of the value that LLMs add to the software design process and justify their integration into the software development workflow. 

\begin{tcolorbox}
\textbf{RQ4: What are the limitations of using LLMs in software design?}
\end{tcolorbox}
\textbf{Rationale}: This RQ explores the limitations that developers perceive from using LLMs in software design. It aims to identify the potential specific challenges, such as the accuracy of design suggestions, the potential for introducing design errors, or the difficulty in aligning LLM-generated design with existing architectural standards. Pinpointing these limitations can help develop solutions or mitigation strategies that enhance the effectiveness and reliability of LLMs in software design.  

\subsection{Mining Study Design}\label{sec:MSDesign}

\subsubsection{Data Collection and Labeling}\label{sec:Data Collection}
We collected shared conversation records between developers and ChatGPT from GitHub, by expanding the dataset of our previous work on how developers interact with ChatGPT in GitHub~\citep{Usage_of_ChatGPT_in_Github}. Developers prefer ChatGPT over other LLMs and LLM applications for its performance~\citep{hou2024llm4se_slr}, and OpenAI introduced the ChatGPT conversation sharing feature in May 2023, which allows ChatGPT users to share their conversations as a link. As a result, we could find a multitude of shared conversations between developers and ChatGPT, some of which were related to software design. By collecting shared conversations from GitHub and then extracting those related to software design, we built the dataset for the mining study. 

Five sources were selected to collect shared conversations: Code, Commits, Issues, Pull Requests and Discussions. For each source category, we searched for the target string ``\textsc{chat.openai.com/share/\{conversation\_id\}}'' in GitHub then retrieved the link and content of the shared ChatGPT conversation. GitHub sets a maximum of 1000 items on its webpage search results and RESTful API query results~\citep{GithubRest2022}. Several strategies were used in this study to retrieve as many shared conversations as possible. For example, we searched for the target string from the \textit{Code} source by programming languages using GitHub's ``\textit{language}'' qualifier, and searched from the \textit{Discussions} source by developing a Web crawler due to the absence of GitHub's Restful API support.

In this study, we collected shared ChatGPT conversations on GitHub from May 2023 to January 2025. 
{\revcolor
The previous study~\citep{Usage_of_ChatGPT_in_Github} collected 2,547 shared conversations from GitHub between May 2023 and June 2024. We repeated the collection procedure from June 2024 to January 2025 and retrieved 670 additional conversations. The present dataset, therefore, contains 3,217 conversations. We relabeled all 3,217 conversations using the criteria defined below and retained 291 design-related conversations for data extraction and coding.
}

After cleaning the data by removing duplicate and non-English search results, we labeled the search results according to whether a conversation between developers and ChatGPT was related to software design. The conversations considered relevant constitute the mining study dataset. 

When labeling the shared conversations, we followed a set of inclusion and exclusion criteria. \rev{A conversation was labeled as related to software design if it satisfied at least one of the inclusion criteria \textbf{I1}$\sim$\textbf{I4} and none of the exclusion criteria \textbf{E1}$\sim$\textbf{E3}.} \rev{The inclusion criteria are:}

{\revcolor\begin{itemize}[leftmargin=2.2em,itemsep=0.15em]
\item[\textbf{I1}] If the term ``design pattern'' or a specific design pattern, such as the blackboard pattern, is mentioned in the conversation, we consider this conversation to be related to software design.
\item[\textbf{I2}] If the conversation discusses the topics related to software architecture, such as architectural patterns (MVC, microservices, layered architecture, event-driven, etc.), architectural components (modularization, service boundaries, API design, etc.), and quality attributes (scalability, maintainability, loose coupling, high cohesion), this conversation is considered relevant.
\item[\textbf{I3}] If the conversation explicitly pertains to the organization of code and modules (such as ``\textit{how to delineate service boundaries}'' or ``\textit{the mechanisms of inter-module communication}'') or describes the dependencies between components (such as ``\textit{how to decouple these two subsystems}''), this conversation is considered relevant.
\item[\textbf{I4}] If the conversation pertains to the trade-offs in design decisions, such as comparing the merits and drawbacks of various design options (for instance, ``\textit{considering read and write frequency when selecting a caching strategy}'') or analyzing the impact of design on non-functional requirements (such as ``\textit{how to enhance system fault tolerance through design}''), this conversation is regarded as relevant.
\end{itemize}}

\rev{The exclusion criteria are:}

{\revcolor\begin{itemize}[leftmargin=2.2em,itemsep=0.15em]
\item[\textbf{E1}] If the conversation solely revolves around the minutiae of implementation details, such as discussing specific code syntax (for instance, ``\textit{how is a Python list comprehension written?}'') or debugging a single function or variable error (such as ``\textit{why does this loop throw a null pointer exception?}''), this conversation is considered irrelevant.
\item[\textbf{E2}] If the conversation solely revolves around business rules, such as the ``\textit{calculation logic for shopping cart discounts}'', without addressing the software design, this conversation is considered irrelevant.
\item[\textbf{E3}] If the developer only inquires about IDE usage, dependency installation commands, and other configuration-related matters, such conversation is considered irrelevant.
\end{itemize}}

When the first author was unsure whether a conversation was related, the second and third authors were invited to discuss until they reached an agreement. As a result, we collected 3,217 shared ChatGPT conversations from GitHub, \rev{including 2,547 conversations collected in our previous work~\citep{Usage_of_ChatGPT_in_Github} (from May 2023 to June 2024) and 670 conversations newly collected in this study (from June 2024 to January 2025),} and 291 out of the 3,217 conversations were classified as related to software design. 
{\revcolor
Table~\ref{T:StatisticsOfMiningStudyStepOne} reports the number of conversations retrieved from each GitHub source and the number labeled as related to software design. Of the 3,217 conversations, 2,926 were labeled as unrelated and excluded from data extraction. We do not report counts for individual exclusion criteria because a single exclusion decision may correspond to multiple criteria. Assigning each excluded conversation to a single criterion after labeling would therefore be arbitrary.
}

\begin{table}[!t]
\centering
\small
\setlength\tabcolsep{4pt}
\renewcommand{\arraystretch}{1.2}
\caption{Statistics of the collected conversations from GitHub}
\label{T:StatisticsOfMiningStudyStepOne}
\begin{tabular}{m{0.25\columnwidth}<{\centering} m{0.20\columnwidth}<{\centering} m{0.42\columnwidth}<{\centering}}
\toprule
\textbf{Source} & \textbf{Total} & \textbf{Software Design Related} \\\midrule
\textbf{Code}          & 1,377 & 91 \\\hline
\textbf{Commits}       & 1,067 & 67 \\\hline
\textbf{Issues}        & 377   & 43 \\\hline
\textbf{Pull Requests} & 325   & 63 \\\hline
\textbf{Discussions}   & 71    & 27 \\\hline
\textbf{Total}         & 3,217 & 291 \\
\bottomrule
\end{tabular}
\end{table}

\subsubsection{Data Extraction and Analysis}\label{sec:Data Extraction and Analysis}
\textbf{Data Extraction}\label{DataExtractionPhase}: In this work, we study how developers utilize ChatGPT to address software design issues and what specific design tasks have been supported by ChatGPT. After collecting the shared conversations between developers and ChatGPT that are related to software design, we obtained a total of 291 shared conversations for data analysis. \rev{We first extracted the data items needed to address RQ1 and RQ2 from these conversations.} The first and fourth authors conducted a pilot data extraction by randomly selecting 30 shared conversations. The two authors extracted the data items listed in Table~\ref{T:DataExtraction} independently. \rev{If disagreements arose, the second and third authors joined discussions with the two initial coders until consensus was reached.} After the pilot data extraction, the first author extracted the data items from all the shared conversations. During this process, any uncertain parts were discussed among the first three authors until they reached a consensus to enhance the correctness of the extracted data. Finally, the first author rechecked all the extracted data to further improve the accuracy of the extracted data. Note that within a shared conversation between developers and ChatGPT, various tasks and topics may be involved; consequently, each shared conversation may extract a differing number of data items.

{\revcolor
We used consensus coding, also called negotiated agreement. The first and fourth authors independently coded a pilot sample of 30 conversations. The first author then extracted and initially coded the full dataset; uncertain or borderline cases were discussed by the first three authors until consensus was reached, and the first author rechecked the extracted data. We did not compute Cohen's kappa or Krippendorff's alpha for the full dataset because it was not independently coded twice, which is required for a chance-corrected agreement coefficient.

This procedure is consistent with the inductive analysis used here. The categories were generated through constant comparison rather than applied from a fixed coding frame, and qualitative-methods research recognizes negotiated agreement as an alternative for interpretive, category-generating analysis~\citep{campbell2013coding,mcdonald2019reliability}. The absence of independent double coding remains a limitation. The replication package provides the conversation content, extracted data items, and assigned codes for each of the 376 task instances~\citep{replpack}.
}

\begin{table*}[htbp]
\centering
\small
\setlength\tabcolsep{1.2pt}
\renewcommand{\arraystretch}{1.2}
\caption{\rev{Data items extracted and the corresponding RQs}}
\label{T:DataExtraction}
\scalebox{0.8}{
\begin{tabular}{
m{0.16\textwidth}<{\centering}
m{0.20\textwidth}<{\centering}
>{\raggedright\arraybackslash}m{0.45\textwidth}
m{0.15\textwidth}<{\centering}
}
\toprule
\textbf{\#} & \textbf{Data Item} & \multicolumn{1}{c}{\textbf{Description}} & \textbf{RQ} \\
\midrule
\textbf{D1} & Task types
& The types of software design tasks addressed by ChatGPT for developers in this dialogue.
& RQ1 \\\hline
\textbf{D2} & Number of rounds
& \rev{The number of dialogue rounds in which developers addressed a single design-related task with ChatGPT.}
& RQ2 \\\hline
\textbf{D3} & Purpose of prompt
& The purpose of initiating a conversation or a new topic by the developer.
& RQ2 \\\hline
\textbf{D4} & Level of design
& The design levels associated with tasks related to software design within the conversation.
& RQ2 \\
\bottomrule
\end{tabular}
}
\end{table*}

\textbf{Data Analysis}\label{DataAnalysisPhase}: For the data analysis of RQ1 (design tasks supported by ChatGPT) and RQ2 (use of ChatGPT in software design), we adopted the Open Coding and Constant Comparison method~\citep{CC2014}, involving three key steps. In the first step, \textit{initial coding}, the first author reviewed the content of each shared ChatGPT conversation related to software design and assigned descriptive codes close to the developers' prompts and ChatGPT's replies. For example, prompts asking ChatGPT to design REST endpoints between a mobile app and a backend, compare a microservices architecture with a modular monolith, define PostgreSQL entities and relationships, or refactor a notification service using the Strategy/Factory patterns were initially coded as ``\textit{API specification design}'', ``\textit{architecture trade-off analysis}'', ``\textit{database schema modeling}'', and ``\textit{pattern-based refactoring}'', respectively. The next step is \textit{focused coding}, where the most frequent codes were compared and grouped into higher-level categories. For RQ1, we consolidated codes such as ``\textit{API specification design}'', ``\textit{protocol selection}'', and ``\textit{interface encapsulation}'' into ``\textit{Interface and Protocol Design}'', and merged ``\textit{database schema modeling}'' and ``\textit{entity relationship definition}'' into ``\textit{Data Model Design}''. For RQ2, we assigned codes such as ``\textit{asking for a design recommendation}'', ``\textit{querying a design concept}'', and ``\textit{generating code under design constraints}'' to the higher categories ``\textit{Recommendation of Design Solution}'', ``\textit{Knowledge Query about Design}'', and ``\textit{Code Generation for Design}'', respectively.

Third, we conducted \textit{theoretical coding} to establish relationships between the identified categories. For example, a prompt asking whether to adopt microservices or a modular monolith for an IoT platform was linked to ``\textit{Architecture Design}'', ``\textit{Recommendation of Design Solution}'', and the ``\textit{Architectural level}'', whereas a prompt asking how to implement an interactive drill-down chart in React and D3 was linked to ``\textit{User Interface Design}'', ``\textit{Code Generation for Design}'', and the ``\textit{Detailed design level}''. For the design-level aspect of RQ2, the categories were derived inductively through the coding process and were subsequently named by following the design levels defined by \citet{PatternOrientedSoftwareArchitecture}. \rev{To reduce individual interpretation bias, disagreements in the derived categories were resolved through collaborative, multi-turn discussions among the first three authors.}

\subsection{Survey Study Design}\label{sec:ESDesign}
In the mining study phase, through RQ1 and RQ2, we have already investigated what kinds of software design tasks developers address by interacting with ChatGPT and the characteristics of their interactions with ChatGPT (e.g., the number of interaction rounds). However, this is still not sufficiently in-depth. We also planned to obtain a deeper understanding of utilizing LLMs for software design from practitioners. Thus, we decided to conduct an industrial survey as a complementary data source to explore \rev{practitioners'} perspectives on the benefits (RQ3) and limitations (RQ4) of employing LLMs for software design, because survey research helps gain information and a deeper understanding of practitioners in order to describe, compare, or explain their knowledge, attitudes, and behavior~\citep{TheSurveyHandbook}. At the same time, the survey study can also \rev{provide a second, practitioner-based perspective on} the findings from our mining study. \rev{We note here, and return to the point in Sections~\ref{sec:RQ1_results} and~\ref{sec:Threats}, that because the answer options of the survey questions for RQ1 and RQ2 were derived from the mining study, the survey complements those findings but cannot independently validate them.} A survey study could be conducted through self-administered questionnaires, telephone surveys, and one-to-one interviews to collect data~\citep{Kitchenham2008}. \rev{After considering the diversity and reliability of the data sources and the efficiency of data collection, we chose an online self-administered questionnaire because it could reach participants in different countries and collect responses at scale~\citep{Kitchenham2008}.}

\subsubsection{Creating the Questionnaire and Recruitment of Participants}\label{sec:Questionnaire Creating and Participants recruitment}
\begin{figure*}[htbp]
\centering
\includegraphics[width=1\textwidth]{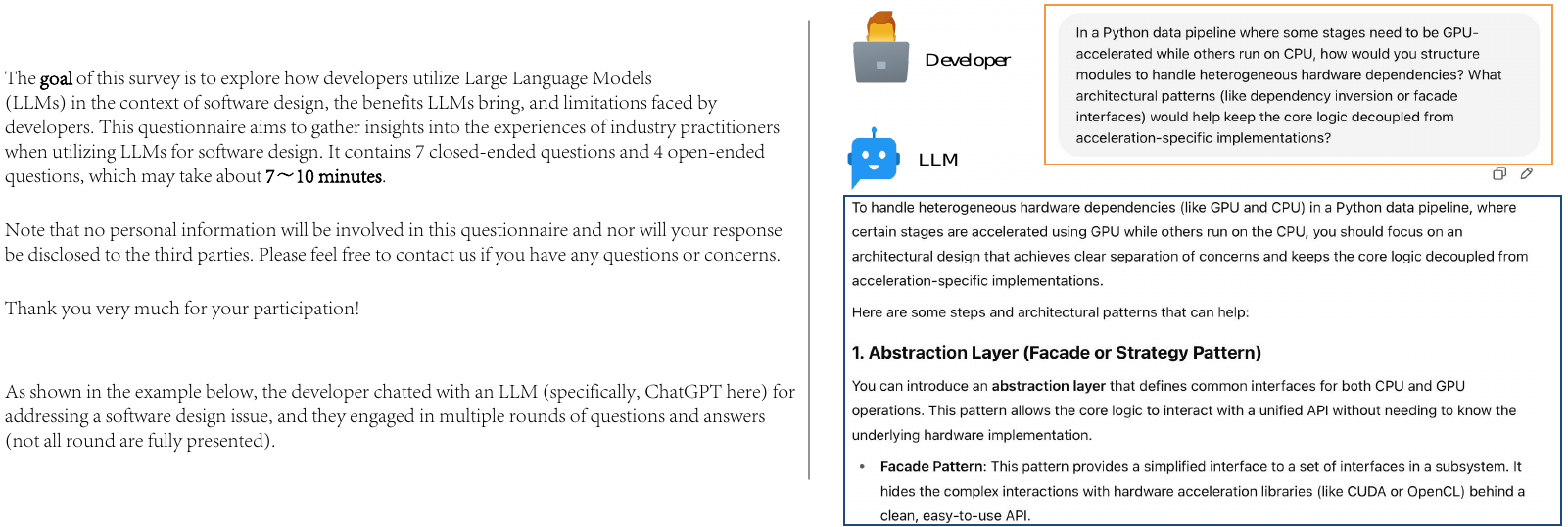}
\caption{Welcome page of the survey questionnaire}
\label{fig:WelcomePageOfTheSurveyQuestionnaire}
\end{figure*}

We formulated the survey questionnaire (see Table~\ref{tab:survey-questions}) that covers all four RQs (RQ1, RQ2, RQ3, and RQ4). The questionnaire, which is in English, \rev{contains a Welcome page followed by 11 questions grouped into five question blocks:} the Welcome page (see Figure~\ref{fig:WelcomePageOfTheSurveyQuestionnaire}) shows the questionnaire requirements, introduction, and an example of a shared conversation related to software design between a developer and ChatGPT; two questions (SQ1 and SQ2) about participants’ background information; one question (SQ3) about participants' frequency of utilizing LLMs to address software design issues; four questions (SQ4, SQ5, SQ6, and SQ7) to \rev{complement} the results of RQ1 and RQ2 from the mining study; three questions (SQ8, SQ9, and SQ10) to answer RQ3 and RQ4; and one question (SQ11) about the willingness to receive the study findings. Among the survey questions for \rev{complementing} RQ1 and RQ2, SQ4 is a semi-open question derived from the results of RQ1, SQ5 and SQ7 are \rev{closed-ended} questions derived from the results of RQ2, while SQ6 is a semi-open question derived from the results of RQ2. \rev{Because these four questions present the categories derived from the mining study as their answer options, and because the option labels were presented to participants without accompanying definitions, they constrain the response space and may anchor participants toward our categorization. We describe how the SQ4 and SQ6 ``Other'' responses were assigned to existing categories in Section~\ref{sec:Results}, and we discuss the anchoring risk in Section~\ref{sec:Threats}.} \rev{The target participants were software practitioners with experience using LLMs to address software development issues}, and we used the following two contact channels to invite potential participants via email:

\begin{itemize}
\item[-] \textbf{Developers from GitHub}: We gathered the email addresses of the developers who shared their conversation records with ChatGPT on GitHub, 
\rev{because these individuals may have experience addressing software design issues with LLMs.}
\item[-] \textbf{Developers from professional software development groups on LinkedIn}: We selected popular professional software development groups on LinkedIn and collected the email addresses of the group members because the members of these groups may use LLMs in their daily development to solve design issues. 
\end{itemize}

Throughout the entire process of the survey study, we adhered to the fundamental ethics of research and the user privacy regulations of relevant platforms (GitHub and LinkedIn). We collected only the email addresses of potential participants to invite them to participate in our survey study via Google Forms. In the questionnaire, we gathered information solely related to participants' experiences in software development, as well as their usage characteristics and attitudes towards LLMs for addressing software design issues. 

\subsubsection{Evaluating and Validating the Questionnaire}\label{sec:validating Questionnaire}
A pilot survey was conducted with the GitHub developers to \rev{assess the comprehensibility and suitability of the survey questionnaire} before formally disseminating survey invitations. We chose to randomly invite 200 developers \rev{who had shared conversations with ChatGPT on GitHub,} from the developers we collected in Section~\ref{sec:Questionnaire Creating and Participants recruitment}. Among the 200 developers, 19 provided valid responses. We analyzed the data from the pilot survey to assess the comprehensibility and efficacy of the survey questions, as well as the overall survey length. The analysis confirmed that the survey questions were unambiguous, the length was suitable for the participant group, and the survey questions successfully elicited meaningful data. Consequently, the questionnaire was deployed for the formal survey without refinement. As the finalized questionnaire was identical to the pilot version, the 19 valid pilot responses were retained and incorporated in the final survey dataset. Eventually, the survey questionnaire contains 4 closed-ended questions (SQ1, SQ3, SQ5 and SQ7), 3 semi-open questions (SQ2, SQ4 and SQ6), and 4 open-ended questions (SQ8, SQ9, SQ10 and SQ11). 

\begin{table*}[htbp]
\centering
\caption{Survey questions on the utilization of LLMs in software design}
\label{tab:survey-questions}
\scalebox{0.8}{
\begin{tabular}{|p{1cm}|p{3.5cm}|p{5.5cm}|p{5.5cm}|}
\hline
\textbf{ID} & \textbf{Type of Questions} & \textbf{Questions} & \textbf{Type of Answers} \\
\hline
SQ1 & Background information about participants & Q1. How many years have you been involved in software development? & \textless{1} year / 1\textasciitilde3 years / 3\textasciitilde5 years / \textgreater{5} years \\
\hline
SQ2 & Background information about participants & Q2. What role do you play in the software development of your organization? & Project manager / Team Leader / Requirements Engineer / Consultant / Architect / Developer / Tester / Other \\
\hline
SQ3 & Participants' frequency of utilizing LLMs to address software design issues & Q3. How often do you chat with LLMs for software design issues? & Never / Very few / Sometimes / Whenever I encounter such issues \\
\hline
SQ4 & \rev{For complementing RQ1 and RQ2} & Q4. Which design tasks have been supported by LLMs according to your experience in software development? & \textit{Interface and Protocol Design} / \textit{Architecture Design} / \textit{Data Model Design} / \textit{Code Refactoring} / \textit{Component Dependency Optimization} / \textit{Performance Optimization} / \textit{Security Design} / \textit{Use of Design Patterns} / \textit{User Interface Design} / Other \\
\hline
SQ5 & \rev{For complementing RQ1 and RQ2} & Q5. How many rounds of dialogue with an LLM are typically required to resolve a design issue or to achieve a satisfactory outcome? & \rev{Single round / 2\textasciitilde3 rounds / 4\textasciitilde10 rounds / More than 10 rounds} \\
\hline
SQ6 & \rev{For complementing RQ1 and RQ2} & Q6. What are the typical purposes of your prompts for addressing software design issues using LLMs? & \textit{Recommendation of Design Solution} / \textit{Knowledge Query about Design} / \textit{Verification of Design Solution} / \textit{Code Generation for Design} / Other \\
\hline
SQ7 & \rev{For complementing RQ1 and RQ2} & Q7. What levels of software design issues can be addressed utilizing LLMs? & \textit{Architectural level} / \textit{Detailed design level} / \textit{Code idiom level} \\
\hline
SQ8 & For answering RQ3 and RQ4 & Q8. What are the benefits of using LLMs in software design according to your experience? & Free text \\
\hline
SQ9 & For answering RQ3 and RQ4 & Q9. What are the limitations and challenges of using LLMs in software design according to your experience? & Free text \\
\hline
SQ10 & For answering RQ3 and RQ4 & Q10. Do you have any further comments about utilization of LLMs for addressing software design issues? & Free text \\
\hline
SQ11 & Will to get the research findings & Q11. Do you want to get the results of this survey? If so, please provide your email address. & Free text \\
\hline
\end{tabular}
}
\end{table*}

\subsubsection{Conducting the Survey and Analyzing Survey Data}\label{sec:Data Conducting surve}
We sent survey invitations to the potential participants via email on 6 June 2025, with a total of 1,130 invitation emails distributed, including 200 pilot invitations. As of 21 June 2025, we had collected 65 valid responses, including 19 from the pilot survey, yielding a response rate of approximately 5.8\%. We also received 4 non-English responses: two in German, one in Japanese, and one in French. After translating them into English, these responses were all regarded as valid responses. We consider responses that exhibit any of the following three characteristics to be invalid: \textbf{(1)} The \rev{participant} provided logically inconsistent answers to open-ended questions. For example, the benefits of LLMs for software design issues mentioned in SQ8 are also cited as drawbacks in SQ9. \textbf{(2)} When answering \rev{SQ3} (``\textit{How often do you \rev{chat} with LLMs for software design issues?}''), the \rev{participants} selected ``\textbf{Never}''. \textbf{(3)} For open-ended questions, the \rev{participants} provided responses that were excessively brief and off-topic. For instance, in response to SQ8 (``\textit{What are the benefits of using LLMs in software design according to your experience?}''), the participants merely replied with ``\textbf{Good}''. For the two distinct groups of survey participants (developers from LinkedIn, developers from GitHub who shared conversations with ChatGPT), we utilized two independent yet content-identical Google Forms links to differentiate the sources of the participants. As a result, among the 65 valid responses, 34 originated from GitHub developers, while 31 came from developers within professional groups on LinkedIn.

We used descriptive statistics~\citep{kaur2018descriptive} \rev{and the Open Coding and Constant Comparison method}~\citep{CC2014} to analyze the answers to the closed-ended questions, semi-open questions, and open-ended questions. Table~\ref{tab:survey_analysis_methods} presents the survey questions and their analysis methods for answering the RQs.

{\revcolor
We analyzed the free-text answers to SQ8, SQ9, and SQ10 using the same open coding and constant comparison procedure. The first author assigned descriptive codes close to the participants' wording and grouped them into categories. The first three authors reviewed the categories against the answers and resolved disagreements through discussion. This process produced seven benefit categories and six limitation categories reported in Section~\ref{sec:RQ3_results} and Section~\ref{sec:RQ4_results}, respectively. 

A single answer could mention several benefits or limitations and could therefore be assigned to several categories. The category counts consequently exceed 65 participants. SQ10 is used to ask for further comments; we coded an answer from SQ10 only when it expresses a benefit or limitation and assigned it to the corresponding category set. Comments that express neither were not coded.
}

\begin{table*}[htbp]
\centering
\caption{Survey questions and according analysis methods to answer the RQs}
\label{tab:survey_analysis_methods}
\scalebox{0.8}{
\begin{tabular}{lll}
\toprule
\textbf{Survey Question} & \textbf{Data Analysis Method}           & \textbf{RQ} \\
\midrule
SQ1 $\sim$ SQ3           & Descriptive Statistics                  & Demographic \\
SQ4                      & Descriptive Statistics                  & RQ1         \\
SQ5 $\sim$ SQ7           & Descriptive Statistics                  & RQ2         \\
SQ8                      & Open Coding and Constant Comparison     & RQ3         \\
SQ9                      & \rev{Open Coding and Constant Comparison} & RQ4         \\
SQ10                     & Open Coding and Constant Comparison     & RQ3 \& RQ4         \\
SQ11                     & No need to analyze                      & Administrative Question         \\
\bottomrule
\end{tabular}
}
\end{table*}

The distributions of software development experience among the survey participants and the roles they play in software development are shown in Figure~\ref{fig:ParticipantDistributions}(a) and Figure~\ref{fig:ParticipantDistributions}(b), respectively. We can see that about 71\% (46 out of 65 participants) possess over three years of development experience, with 45\% (29 out of 65 participants) having more than five years of experience. Moreover, the roles of the participants are widely distributed, with 41\% taking the role of ``Developer'', 18\% opting for ``Tester'', and the remaining 40\% choosing roles such as ``Project Manager'', ``Team Leader'', ``Requirements Engineer'', ``Consultant'', and ``Architect''. Specifically, the Survey Study (SQ8\textasciitilde SQ10) provides responses for answering RQ3 and RQ4, while the answers for RQ1 and RQ2 from the Mining Study are complemented by the relevant survey questions (SQ4\textasciitilde SQ7).

\definecolor{expunderone}{HTML}{D1495B}
\definecolor{exponeThree}{HTML}{EE964B}
\definecolor{expthreefive}{HTML}{F4D35E}
\definecolor{expfiveplus}{HTML}{00A6A6}
\definecolor{roledeveloper}{HTML}{5B5F97}
\definecolor{roletester}{HTML}{FFC145}
\definecolor{rolearchitect}{HTML}{2D936C}
\definecolor{rolerequirements}{HTML}{D8572A}
\definecolor{roleproject}{HTML}{00A7E1}
\definecolor{roleteam}{HTML}{A53860}
\definecolor{roleconsultant}{HTML}{7CB518}

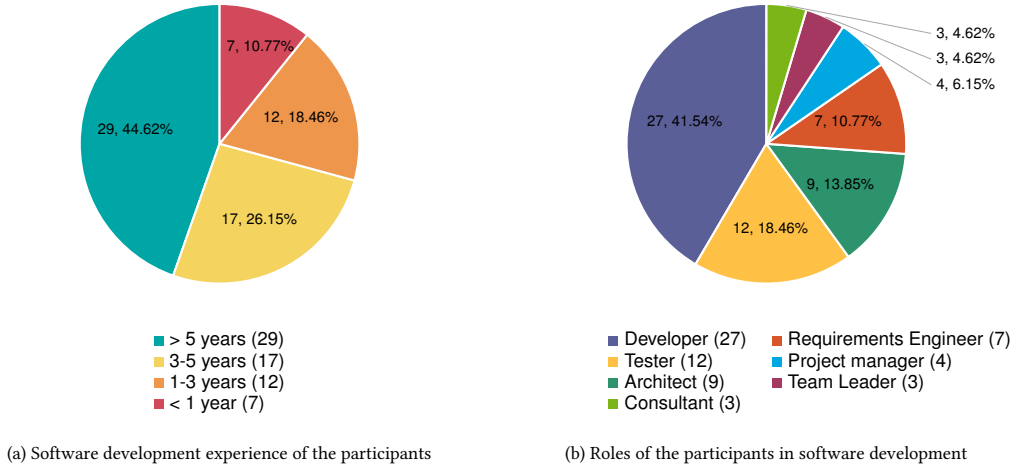
\begin{figure*}[htbp]
\centering
\begin{minipage}[t]{0.48\linewidth}
\centering
\begin{tikzpicture}[font=\piechartfont\scriptsize, line join=round]
\def\pieRadius{1.85}
\path[use as bounding box] (-2.70,-2.05) rectangle (2.70,2.05);
\pgfmathsetmacro{\currentAngle}{90}
\foreach \name/\slicevalue/\percentlabel/\slicecolor in {
  {> 5 years}/29/44.62/expfiveplus,
  {3-5 years}/17/26.15/expthreefive,
  {1-3 years}/12/18.46/exponeThree,
  {< 1 year}/7/10.77/expunderone%
}{
  \pgfmathsetmacro{\deltaAngle}{\slicevalue/65*360}
  \pgfmathsetmacro{\nextAngle}{\currentAngle+\deltaAngle}
  \pgfmathsetmacro{\midAngle}{\currentAngle+0.5*\deltaAngle}
  \filldraw[fill=\slicecolor, draw=white, line width=0.8pt]
    (0,0) -- (\currentAngle:\pieRadius) arc (\currentAngle:\nextAngle:\pieRadius) -- cycle;
  \ifnum\slicevalue=7
    \node[font=\pieslicelabelfont, inner sep=0pt] at ([xshift=4.8pt,yshift=6.6pt]\midAngle:1.12)
      {\pielabeltext{\slicevalue}{\percentlabel}};
  \else
    \node[font=\pieslicelabelfont, inner sep=0pt] at (\midAngle:1.13)
      {\pielabeltext{\slicevalue}{\percentlabel}};
  \fi
  \xdef\currentAngle{\nextAngle}
}
\end{tikzpicture}

\vspace{0.3em}
\begin{minipage}[c][4.8em][c]{\linewidth}
\centering
{\tikzlegendfont
\begin{tabular}{@{}l@{}}
\textcolor{expfiveplus}{\rule{1.0ex}{1.0ex}}~> 5 years (29) \\
\textcolor{expthreefive}{\rule{1.0ex}{1.0ex}}~3-5 years (17) \\
\textcolor{exponeThree}{\rule{1.0ex}{1.0ex}}~1-3 years (12) \\
\textcolor{expunderone}{\rule{1.0ex}{1.0ex}}~< 1 year (7)
\end{tabular}}
\end{minipage}

\vspace{0.25em}
{\scriptsize\normalfont (a) Software development experience of the participants\par}
\end{minipage}\hfill
\begin{minipage}[t]{0.48\linewidth}
\centering
\begin{tikzpicture}[font=\piechartfont\scriptsize, line join=round]
\def\pieRadius{1.85}
\path[use as bounding box] (-3.20,-2.05) rectangle (3.20,2.05);
\pgfmathsetmacro{\currentAngle}{90}
\foreach \name/\slicevalue/\percentlabel/\slicecolor/\externalY in {
  {Developer}/27/41.54/roledeveloper/0,
  {Tester}/12/18.46/roletester/0,
  {Architect}/9/13.85/rolearchitect/0,
  {Requirements Engineer}/7/10.77/rolerequirements/0,
  {Project manager}/4/6.15/roleproject/0.78,
  {Team Leader}/3/4.62/roleteam/1.12,
  {Consultant}/3/4.62/roleconsultant/1.46%
}{
  \pgfmathsetmacro{\deltaAngle}{\slicevalue/65*360}
  \pgfmathsetmacro{\nextAngle}{\currentAngle+\deltaAngle}
  \pgfmathsetmacro{\midAngle}{\currentAngle+0.5*\deltaAngle}
  \filldraw[fill=\slicecolor, draw=white, line width=0.8pt]
    (0,0) -- (\currentAngle:\pieRadius) arc (\currentAngle:\nextAngle:\pieRadius) -- cycle;
  \ifnum\slicevalue<5
    \draw[gray!70, line width=0.35pt, overlay] (\midAngle:\pieRadius) -- (2.18,\externalY);
    \node[font=\pieslicelabelfont, anchor=west, inner sep=0.6pt, overlay] at (2.22,\externalY)
      {\pielabeltext{\slicevalue}{\percentlabel}};
  \else
    \node[font=\pieslicelabelfont, inner sep=0pt] at (\midAngle:1.12)
      {\pielabeltext{\slicevalue}{\percentlabel}};
  \fi
  \xdef\currentAngle{\nextAngle}
}
\end{tikzpicture}

\vspace{0.3em}
\begin{minipage}[c][4.8em][c]{\linewidth}
\centering
{\tikzlegendfont
\makebox[6.58cm][c]{%
\begin{tabular}{@{}c@{}}
\makebox[3.22cm][l]{\hspace{1.2cm}\textcolor{roledeveloper}{\rule{1.0ex}{1.0ex}}\hspace{0.35em}Developer (27)}\hspace{0.14cm}%
\makebox[3.22cm][l]{\textcolor{rolerequirements}{\rule{1.0ex}{1.0ex}}\hspace{0.35em}Requirements Engineer (7)} \\
\makebox[3.22cm][l]{\hspace{1.2cm}\textcolor{roletester}{\rule{1.0ex}{1.0ex}}\hspace{0.35em}Tester (12)}\hspace{0.14cm}%
\makebox[3.22cm][l]{\textcolor{roleproject}{\rule{1.0ex}{1.0ex}}\hspace{0.35em}Project manager (4)} \\
\makebox[3.22cm][l]{\hspace{1.2cm}\textcolor{rolearchitect}{\rule{1.0ex}{1.0ex}}\hspace{0.35em}Architect (9)}\hspace{0.14cm}%
\makebox[3.22cm][l]{\textcolor{roleteam}{\rule{1.0ex}{1.0ex}}\hspace{0.35em}Team Leader (3)} \\
\makebox[3.22cm][l]{\hspace{1.2cm}\textcolor{roleconsultant}{\rule{1.0ex}{1.0ex}}\hspace{0.35em}Consultant (3)}\hspace{0.14cm}%
\makebox[3.22cm][l]{}
\end{tabular}}}
\end{minipage}

\vspace{0.25em}
{\scriptsize\normalfont (b) Roles of the participants in software development\par}
\end{minipage}
\caption{Distributions of software development experience and roles of the participants}
\label{fig:ParticipantDistributions}
\label{F:ExperienceDistribution}
\label{F:RoleDistribution}
\end{figure*}

\section{Results}\label{sec:Results}
In this section, we present the results of our mixed-methods study that answers the four RQs defined in Section~\ref{sec:RQs}.

\rev{One interpretive convention applies to every frequency reported below. The counts and percentages from the mining study describe the 291 shared conversations and the 376 design task instances extracted from them, and the counts and percentages from the survey describe the 65 participants who responded. Neither set is an estimate for a wider population: the conversations were publicly shared by developers who chose to share them, and the participants are a self-selected subset of those we invited (see Section~\ref{sec:Threats}). Statements such as ``the most frequent category'' therefore mean ``the most frequent in this dataset'' throughout, and we used no inferential statistics to compare the two data sources.}

\subsection{RQ1: What specific software design tasks have been supported by ChatGPT?}\label{sec:RQ1_results}

\subsubsection{Mining Study Results}
\begin{figure*}[htbp]
\centering
\makebox[\textwidth][c]{%
  \hspace*{-0.05\textwidth}%
  \includegraphics[width=0.9\textwidth]{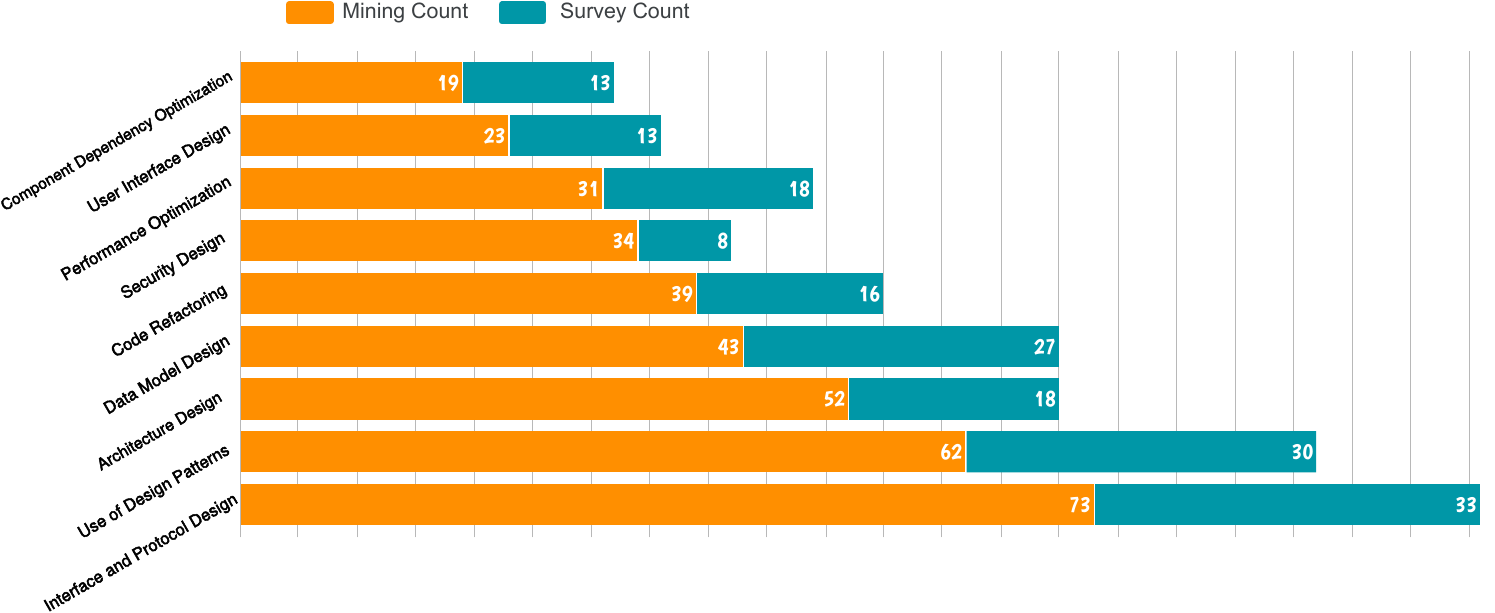}%
}
\caption{Counts of the nine software design task categories. \rev{The left part (\textcolor{orange}{orange}) counts task instances extracted in the mining study ($n$~=~376 task instances); the right part (\textcolor{Cyan}{cyan}) counts survey participants who selected the category in SQ4 ($n$~=~65 participants, multiple-response). The two sides therefore use different units of analysis and are not directly comparable as counts.}}
\label{fig:The quantities of the nine task categories}
\end{figure*}

To answer this RQ, we used the Open Coding and Constant Comparison method~\citep{CC2014} to identify the software design tasks supported by ChatGPT in the mined conversations. We finally derived nine primary categories of software design tasks: \textit{Interface and Protocol Design}, \textit{Architecture Design}, \textit{Data Model Design}, \textit{Code Refactoring}, \textit{Component Dependency Optimization}, \textit{Performance Optimization}, \textit{Security Design}, \textit{Use of Design Patterns}, and \textit{User Interface Design}. In the dataset of 291 shared conversations, we identified 376 software design tasks across the nine categories. Figure~\ref{fig:The quantities of the nine task categories} shows the distribution of the design task categories in the mining study on the left side of the mixed bar chart in \textcolor{orange}{orange}. Note that one shared conversation can include up to six tasks belonging to four design task categories; however, most conversations involve one task, often with multiple dialogue rounds (see Section~\ref{sec:RQ2_results}). Below, we describe each task category and provide one representative example, including its ChatGPT share link and the brief description of the full conversation.

{\revcolor
We applied three assignment rules. The unit of assignment is the \textit{task instance} rather than the conversation, so a conversation with several design tasks contributes several task instances. Each task instance receives one primary category, determined by the developer's stated design goal. Cases that could plausibly receive more than one category were discussed by the first three authors, as described in Section~\ref{sec:Data Extraction and Analysis}. Table~\ref{T:CategoryDefinitions} provides the operational definitions of nine design task categories and their boundary rules. Table~\ref{T:SubCodes} reports the sub-codes grouped into each category and their counts. 

}

{\footnotesize\revcolor
\begin{table*}[htbp]
\centering
\caption{\rev{Operational definitions of the nine design task categories and the boundary rules applied when a task instance could be assigned to more than one category}}
\label{T:CategoryDefinitions}
\begin{tabular}{p{2.8cm} p{7.1cm} p{6.3cm}}
\toprule
\textbf{Category} & \textbf{Assigned when the developer's stated design goal is} & \textbf{Assigned elsewhere when} \\
\midrule
\textit{Interface and Protocol Design} & to specify a contract for interaction between parts of the system or with external systems, e.g., API endpoints and payloads, the choice of a communication protocol, or the encapsulation of an existing interface. & the interface is discussed only as one element of a system-wide decomposition or style decision ($\rightarrow$ \textit{Architecture Design}), or the request concerns only the syntax of a client call (excluded by \textbf{E1}). \\
\midrule
\textit{Architecture Design} & to determine the system-wide structure, i.e., decomposition into modules or services, the choice of an architectural style, the coordination among components, or the trade-offs among such alternatives. & the task narrows to a single interaction contract ($\rightarrow$ \textit{Interface and Protocol Design}), to the dependency structure among components that already exist ($\rightarrow$ \textit{Component Dependency Optimization}), or to a single quality attribute pursued through tactics ($\rightarrow$ \textit{Performance Optimization}, \textit{Security Design}). \\
\midrule
\textit{Data Model Design} & to structure the data, i.e., entities and their relationships, database schemas, data flow, or data mapping strategies. & the schema is a byproduct of deciding the system decomposition ($\rightarrow$ \textit{Architecture Design}), or the request concerns query syntax (excluded by \textbf{E1}). \\
\midrule
\textit{Code Refactoring} & to restructure existing code while preserving its behavior, e.g., extracting reusable logic, decoupling modules, or improving an existing structure. & the developer asks which named pattern to adopt ($\rightarrow$ \textit{Use of Design Patterns}), the target is the dependency structure among components ($\rightarrow$ \textit{Component Dependency Optimization}), or the target is a performance property ($\rightarrow$ \textit{Performance Optimization}). \\
\midrule
\textit{Component Dependency Optimization} & to improve the dependency structure among existing components, e.g., resolving circular dependencies, reducing coupling, organizing dependency injection, or isolating external dependencies. & the developer is deciding the decomposition or the architectural style itself rather than repairing the dependencies of an existing one ($\rightarrow$ \textit{Architecture Design}). \\
\midrule
\textit{Performance Optimization} & to achieve a performance property such as latency, throughput, or resource usage through design-level means, e.g., caching, concurrency, or performance monitoring. & performance appears only as one criterion within a broader design trade-off ($\rightarrow$ \textit{Architecture Design}), or the request is a local code-level micro-optimization (excluded by \textbf{E1}). \\
\midrule
\textit{Security Design} & to achieve a security property through design-level mechanisms, e.g., authentication, authorization, encryption, or threat modeling. & the request concerns a specific vulnerability in one code fragment rather than a design-level mechanism (excluded by \textbf{E1}). \\
\midrule
\textit{Use of Design Patterns} & to select, compare, or apply a named design pattern. When a task both selects a pattern and applies it to existing code, it is assigned here if the developer's question is \textit{which} pattern to use. & the developer's question is \textit{how} to restructure the code and the pattern is only the means of doing so ($\rightarrow$ \textit{Code Refactoring}). \\
\midrule
\textit{User Interface Design} & to design the front end, i.e., component layout, interaction logic, client-side state management, or user experience. & the task concerns the interface that the front end consumes rather than the front end itself ($\rightarrow$ \textit{Interface and Protocol Design}). \\
\bottomrule
\end{tabular}
\end{table*}
}

{\footnotesize\revcolor
\begin{table*}[htbp]
\centering
\caption{\rev{Sub-codes identified within each of the nine design task categories, with the number of task instances assigned to each sub-code ($n$~=~376 task instances) and sub-code counts sum to the category total}}
\label{T:SubCodes}
\begin{tabular}{p{3.0cm} p{0.8cm} p{11.5cm}}
\toprule
\textbf{Category} & \textbf{$n$} & \textbf{Sub-codes (count)} \\
\midrule
\textit{Interface and Protocol Design} & 73 & Endpoint and boundary specification (37); Protocol selection and comparison (11); Interface encapsulation of existing components (10); Event payload and schema evolution (8); API versioning and backward compatibility (7) \\
\midrule
\textit{Use of Design Patterns} & 62 & Behavioral patterns, i.e., strategy, state, command, template method, chain of responsibility, mediator, and observer (43); Structural patterns, i.e., adapter and decorator (10); Creational patterns, i.e., factory (9) \\
\midrule
\textit{Architecture Design} & 52 & System decomposition and modularization (17); Event-driven and messaging integration (10); Stream-processing and pipeline topology (9); Workflow orchestration boundaries (7); Multi-tenancy and tenant isolation (6); Read/write separation and CQRS (3) \\
\midrule
\textit{Data Model Design} & 43 & Domain entity and schema design (24); Access control and ownership modeling (12); Soft deletion and archival (4); Audit trail and revision history (3) \\
\midrule
\textit{Code Refactoring} & 39 & Cross-cutting concern extraction (15); Conditional logic decomposition (13); Responsibility and boundary cleanup (11) \\
\midrule
\textit{Security Design} & 34 & Upload integrity and API abuse mitigation (10); Authentication and session hardening (9); Authorization, permission boundaries, and threat modeling (8); Sensitive data protection (7) \\
\midrule
\textit{Performance Optimization} & 31 & Pre-aggregation and read-path optimization (9); Caching and cache invalidation (8); Latency reduction (8); Throughput and batching (6) \\
\midrule
\textit{User Interface Design} & 23 & Interaction and state models (13); Multi-step flow and wizard design (8); Component hierarchy and data presentation (2) \\
\midrule
\textit{Component Dependency Optimization} & 19 & Boundary hardening and module isolation (10); Dependency cycle and injection untangling (5); Fan-out reduction (4) \\
\bottomrule
\end{tabular}
\end{table*}
}

\textbf{(1) \textit{Interface and Protocol Design}} is the most frequent category in the mining study, covering 73 of the 376 identified design tasks (19.4\%). It refers to defining the specifications for internal and external interactions of  the system, including API design, protocol implementation, and interface encapsulation and optimization.
\begin{tcolorbox}[colback=black!0, colframe=black!20, width=1.0\linewidth, arc=0.5mm, auto outer arc, boxrule=0.8pt, fontupper=\small]
{\textbf{ChatGPT Share Link}: \url{https://chatgpt.com/share/69eaf54f-7a50-83e8-9978-8b16702a14f1}\\
\textbf{Description}: This conversation demonstrates a developer leveraging an LLM to \rev{design the API between a mobile app and its backend for a new software project}. The developer begins with a high-level concept for a ``Smart Pantry'' system. The LLM helps refine this idea into a concrete technical plan by proposing a standard: It suggests using a RESTful API, a common best practice for mobile-to-backend communication.
}\end{tcolorbox}

\textbf{(2) \textit{Architecture Design}} covers 52 identified design tasks (13.8\%). It refers to high-level design activities, including system module decomposition, organizational structure design, architectural trade-offs, and selection of interaction and coordination patterns among components.
\begin{tcolorbox}[colback=black!0, colframe=black!20, width=1.0\linewidth, arc=0.5mm, auto outer arc, boxrule=0.8pt, fontupper=\small]
{\textbf{ChatGPT Share Link}: \url{https://chatgpt.com/share/69eaf9f7-e0bc-83e8-b3dc-566cb03d0376}\\
\textbf{Description}: \rev{A developer seeks assistance in designing an architecture for an agricultural IoT platform. The LLM proposes a multi-tenant system with MQTT/Kafka ingestion, stream processing, time-series storage, and real-time dashboards. It suggests implementing the alert engine with compiled, stateful stream rules (Flink/Kafka Streams) using windowing and deduplication, and adding an external materialized state layer to make the stream state inspectable for observability and debugging.}
}\end{tcolorbox}

\textbf{(3) \textit{Data Model Design}} includes 43 identified design tasks (11.4\%). It refers to modeling data entities and their relationships, and some other related tasks like design of data flow, database structure, and data mapping strategies.
\begin{tcolorbox}[colback=black!0, colframe=black!20, width=1.0\linewidth, arc=0.5mm, auto outer arc, boxrule=0.8pt, fontupper=\small]
{\textbf{ChatGPT Share Link}: \url{https://chatgpt.com/share/69eafc2f-8e78-83e8-8cf7-c56b8b5774a1}\\
\textbf{Description}: A software developer asks the LLM to design a relational database schema for a new community management application. The LLM provides a complete set of PostgreSQL DDL (Data Definition Language) statements for the required features and follows up with clarifying questions to address potential future requirements and edge cases.
}\end{tcolorbox}

\textbf{(4) \textit{Code Refactoring}} contains 39 identified design tasks (10.4\%). It refers to refactoring code structure by \rev{applying strategies such as logical reuse, module decoupling, and responsibility cleanup.}
\begin{tcolorbox}[colback=black!0, colframe=black!20, width=1.0\linewidth, arc=0.5mm, auto outer arc, boxrule=0.8pt, fontupper=\small]
{\textbf{ChatGPT Share Link}: \url{https://chatgpt.com/share/69eafd76-9cb8-83e8-9f7c-3799ecdaba01}\\
\textbf{Description}: A software developer asks the LLM for advice on refactoring a class with a large conditional block. The LLM identifies the violation of the Open/Closed Principle and provides a detailed solution using the Strategy and Factory design patterns, including conceptual code examples.
}\end{tcolorbox}

\textbf{(5) \textit{Component Dependency Optimization}} covers 19 identified design tasks (5.1\%). It refers to tasks that specifically aim to improve the dependency structure among components, such as resolving circular dependencies, reducing excessive coupling, optimizing dependency injection, and isolating external dependencies.
\begin{tcolorbox}[colback=black!0, colframe=black!20, width=1.0\linewidth, arc=0.5mm, auto outer arc, boxrule=0.8pt, fontupper=\small]
{\textbf{ChatGPT Share Link}: \url{https://chatgpt.com/share/69eafe47-592c-83e8-8d39-220551304035}\\
\textbf{Description}: A software developer asks for high-level design strategies to resolve circular and fan-out dependencies in their microservices project. The LLM responds by explaining how to apply design patterns like Event-Driven Architecture and the Mediator Pattern to decouple the services.
}\end{tcolorbox}

\textbf{(6) \textit{Performance Optimization}} accounts for 31 identified design tasks (8.2\%). It refers to improving system performance through multiple architectural tactics like performance monitoring, caching strategies, and concurrent processing.
\begin{tcolorbox}[colback=black!0, colframe=black!20, width=1.0\linewidth, arc=0.5mm, auto outer arc, boxrule=0.8pt, fontupper=\small]
{\textbf{ChatGPT Share Link}: \url{https://chatgpt.com/share/69eafe47-592c-83e8-p0s9-bbyl7yusam2f}\\
\textbf{Description}: A software developer seeks design-level advice from an LLM to optimize a slow analytics dashboard. The LLM provides initial strategies focusing on separating analytical workloads from transactional ones, specifically through asynchronous pre-aggregation, caching, and database read replicas.
}\end{tcolorbox}

\textbf{(7) \textit{Security Design}} contains 34 identified design tasks (9.0\%). It refers to applying multiple mechanisms to ensure system security, such as authentication and encryption.
\begin{tcolorbox}[colback=black!0, colframe=black!20, width=1.0\linewidth, arc=0.5mm, auto outer arc, boxrule=0.8pt, fontupper=\small]
{\textbf{ChatGPT Share Link}: \url{https://chatgpt.com/share/69eaff83-70e8-83e8-be15-232efe7e86f3}\\
\textbf{Description}: A software developer asks an LLM for help with the security design and threat modeling of a new user profile microservice. The LLM proposes using the STRIDE threat modeling framework and asks for more specific details about the service's authentication, data flow, and architecture to begin the security analysis.
}\end{tcolorbox}

\textbf{(8) \textit{Use of Design Patterns}} is the second most frequent category, covering 62 identified design tasks (16.5\%). It refers to applying concrete design patterns (such as the Factory design pattern) to achieve the best practices of detailed design.
\begin{tcolorbox}[colback=black!0, colframe=black!20, width=1.0\linewidth, arc=0.5mm, auto outer arc, boxrule=0.8pt, fontupper=\small]
{\textbf{ChatGPT Share Link}: \url{https://chatgpt.com/share/69eb0022-a2ac-83e8-b5a3-75bfcec82787}\\
\textbf{Description}: A software developer asks for help refactoring a notification service that uses a large \texttt{if-else} statement. The LLM responds with a detailed solution, recommending a combination of the Strategy and Factory design patterns and providing complete, refactored Java code examples for implementation.
}\end{tcolorbox}

\textbf{(9) \textit{User Interface Design}} includes 23 identified design tasks (6.1\%). It refers to front-end related design activities, including design of front-end component layout, implementation of interactive logic, state management, and optimization of user experience.
\begin{tcolorbox}[colback=black!0, colframe=black!20, width=1.0\linewidth, arc=0.5mm, auto outer arc, boxrule=0.8pt, fontupper=\small]
{\textbf{ChatGPT Share Link}: \url{https://chatgpt.com/share/69eb0091-4678-83e8-a3b6-f5e6c2337500}\\
\textbf{Description}: A software developer asks an LLM how to implement an interactive, animated ``drill-down'' feature on a D3.js chart within a React application. The LLM provides a comprehensive solution, including a strategy for state management using React hooks and a detailed code snippet in TypeScript demonstrating how to handle the enter/exit D3 transitions to create a smooth animation between data views.
}\end{tcolorbox}

\subsubsection{Survey Study Results}
The survey question (SQ4) in the questionnaire \rev{lists the nine design task categories introduced above, derived from the mining study as candidate options}.  \rev{Participants could select multiple task categories in this question and could also fill in} the “Other” field to express other design tasks which have been supported by LLMs and are not covered by the candidate options. In the 65 responses, 7 participants completed the “Other” field. Following the authors' discussion, these 7 responses can be categorized into the existing nine types of tasks (3 into \textit{Use of Design Patterns}, 2 into \textit{Architecture Design}, and 2 into \textit{Data Model Design}), with no new categories identified. The right part of Figure~\ref{fig:The quantities of the nine task categories}, \rev{shown in \textcolor{cyan}{cyan}, presents} the distribution of the design tasks selected by the survey participants. A total of 176 \rev{task-category selections were made} by the participants. Two participants selected all nine design task categories, three chose eight, and on average, each participant selected 2.7 task categories. Although the left and right parts of Figure~\ref{fig:The quantities of the nine task categories} present data in a similar format, they convey different messages. The left part in \textcolor{orange}{orange} presents the number of tasks identified by the authors from the dataset of shared conversations, while the right part in \textcolor{cyan}{cyan} shows the number of the design task categories selected by the survey participants based on their experiences. \rev{The units of analysis therefore differ: the mining study counts 376 task instances, of which each falls into exactly one category, whereas the survey counts, for each category, how many of the 65 participants selected it, and a participant may select several categories. For this reason we do not compare the two sides as counts and do not test them against each other; we only observe that all nine categories were selected by at least some participants and that no participant contributed a task type that fell outside the nine categories.}

{\revcolor
\textbf{Synthesis of the mining and survey results.} The survey complements the mining study because SQ4 presents its nine categories as candidate options. All nine categories were selected by at least one participant, and the seven SQ4 ``Other'' responses were assigned to existing categories. These observations do not establish that the categories are complete or mutually exclusive in other data. We discuss this limitation in Section~\ref{sec:Threats}.
}

\begin{tcolorbox}[breakable, colback=black!5, colframe=black!20, width=1.0\linewidth, arc=1mm, auto outer arc, boxrule=1.5pt]
{\textbf{RQ1 Summary}: \textit{We identified nine categories of software design tasks supported by ChatGPT. The mining study of 291 developer-ChatGPT conversations shows that \rev{the developers who shared these conversations} used ChatGPT for a wide range of software design tasks, from Interface and Protocol Design to Architecture Design, Data Model Design, and User Interface Design. \rev{Each of the nine categories was also selected by at least some of the 65 survey participants as a design task they had addressed with LLMs, and the 7 answers entered under ``Other'' did not require a new category. Because the categories were presented to participants as the answer options of SQ4, this survey evidence complements the mining study results rather than independently confirming the taxonomy.}}}
\end{tcolorbox}

\subsection{RQ2: How \rev{is ChatGPT} used in software design?}\label{sec:RQ2_results}

\subsubsection{Mining Study Results}
RQ2 focuses on the behaviors and intentions of developers when utilizing ChatGPT to address software design issues. During the data analysis phase (see Section~\ref{DataAnalysisPhase}), we identified three aspects that answer this RQ. Below, we provide the descriptions of the three aspects, along with their corresponding analysis methods and results.

\textbf{1) How many rounds of dialogues did the developers engage in with ChatGPT regarding a single design-related task?}

\rev{The number of dialogue rounds provides a descriptive measure of how developers interacted with ChatGPT for a single task, although it does not measure ChatGPT capability or task success.} We employed a statistical approach to answer this aspect by analyzing data item D2, specifically the data item concerning the number of rounds. Figure~\ref{fig:StatisticsOfNumberOfRounds} presents the statistical data regarding the number of dialogue rounds\rev{; the highest number of rounds recorded is 68}. \rev{The distribution spans single-round exchanges and much longer multi-turn interactions.}

{\revcolor
The distribution is right-skewed. Across the 376 task instances, the median is 5 dialogue rounds (IQR 2$\sim$11), the mean is 6.18 (SD 5.33), and the maximum is 68 rounds. We therefore use the median and IQR as the primary measures of interaction length.
}

\definecolor{roundsingle}{HTML}{845EC2}
\definecolor{roundtwothree}{HTML}{FF9671}
\definecolor{roundfourten}{HTML}{FFC75F}
\definecolor{roundoverten}{HTML}{00C9A7}

\begin{figure*}[htbp]
\centering
\begin{tikzpicture}[font=\piechartfont\small, line join=round]
\def\pieRadius{1.85}
\path[use as bounding box] (-6.0,-2.45) rectangle (6.0,2.45);
\pgfmathsetmacro{\currentAngle}{90}
\foreach \name/\slicevalue/\percentlabel/\chartcolor in {
  {Over 10 Rounds}/127/33.78/roundoverten,
  {2-3 Rounds}/106/28.19/roundtwothree,
  {4-10 Rounds}/85/22.61/roundfourten,
  {Single Round}/58/15.43/roundsingle%
}{
  \pgfmathsetmacro{\deltaAngle}{\slicevalue/376*360}
  \pgfmathsetmacro{\nextAngle}{\currentAngle+\deltaAngle}
  \pgfmathsetmacro{\midAngle}{\currentAngle+0.5*\deltaAngle}
  \filldraw[fill=\chartcolor, draw=white, line width=0.9pt]
    (0,0) -- (\currentAngle:\pieRadius) arc (\currentAngle:\nextAngle:\pieRadius) -- cycle;
  \node[font=\pieslicelabelfont] at (\midAngle:1.12)
    {\pielabeltext{\slicevalue}{\percentlabel}};
  \xdef\currentAngle{\nextAngle}
}
\node[anchor=west, align=left, font=\tikzlegendfont, inner sep=0pt, outer sep=0pt]
  at (2.8,0)
  {
\begin{tabular}{@{}l@{}}
\textcolor{roundoverten}{\rule{1.0ex}{1.0ex}}~Over 10 Rounds (127) \\
\textcolor{roundtwothree}{\rule{1.0ex}{1.0ex}}~2-3 Rounds (106) \\
\textcolor{roundfourten}{\rule{1.0ex}{1.0ex}}~4-10 Rounds (85) \\
\textcolor{roundsingle}{\rule{1.0ex}{1.0ex}}~Single Round (58)
    \end{tabular}
  };
\end{tikzpicture}
\caption{Distribution of dialogue rounds per design-related task}
\label{fig:StatisticsOfNumberOfRounds}
\end{figure*}
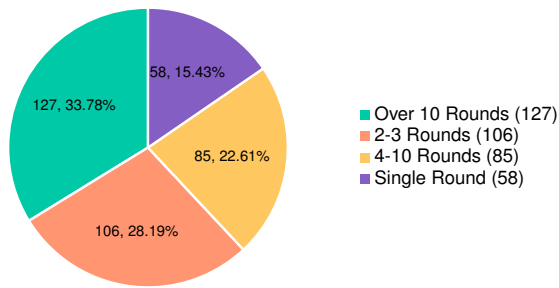

{\revcolor
Table~\ref{tab:RoundsByCategory} shows that interaction length differs across task categories. \textit{Architecture Design} has a median of 11 rounds, no single-round task instances, and 57.7\% of instances with more than 10 rounds. \textit{Use of Design Patterns} has a median of 2 rounds; 48.4\% of its instances contain a single dialogue round, and 3.2\% exceed 10 rounds. \textit{Interface and Protocol Design} has a median of 3 rounds. \textit{Performance Optimization} and \textit{Component Dependency Optimization} have medians of 11 and 12 rounds, respectively; the latter category contains only 19 instances.

The cross-tabulations also show differences in prompt purpose and design level. Architectural-level tasks are most common in \textit{Component Dependency Optimization} (63.2\%), while \textit{Use of Design Patterns} is mainly at the Detailed design level (75.8\%). Knowledge Query about Design is the most common purpose for \textit{Architecture Design} (50.0\%), whereas Recommendation of Design Solution is the most common purpose for \textit{Component Dependency Optimization} (42.1\%). These are descriptive observations; no significance tests were applied.
}

\begin{table*}[htbp]
\revcolor
\centering
\small
\setlength\tabcolsep{4pt}
\renewcommand{\arraystretch}{1.15}
\caption{\rev{Number of dialogue rounds by design task category ($n$~=~376 task instances. Percentages are row percentages within a category. The rightmost column gives the median number of rounds for the category. These are descriptive breakdowns; no significance test is applied.)}}
\label{tab:RoundsByCategory}
\scalebox{0.82}{
\begin{tabular}{lrrrrrr}
\toprule
\textbf{Task category} & \textbf{$n$} & \textbf{Single} & \textbf{2$\sim$3} & \textbf{4$\sim$10} & \textbf{$>$10} & \textbf{Median} \\
\midrule
\textit{Interface and Protocol Design}     & 73  & 5 (6.8\%)   & 37 (50.7\%) & 16 (21.9\%) & 15 (20.5\%) & 3    \\
\textit{Use of Design Patterns}            & 62  & 30 (48.4\%) & 27 (43.5\%) & 3 (4.8\%)   & 2 (3.2\%)   & 2    \\
\textit{Architecture Design}               & 52  & 0 (0.0\%)   & 2 (3.8\%)   & 20 (38.5\%) & 30 (57.7\%) & 11   \\
\textit{Data Model Design}                 & 43  & 7 (16.3\%)  & 14 (32.6\%) & 7 (16.3\%)  & 15 (34.9\%) & 4    \\
\textit{Code Refactoring}                  & 39  & 7 (17.9\%)  & 9 (23.1\%)  & 12 (30.8\%) & 11 (28.2\%) & 5    \\
\textit{Security Design}                   & 34  & 0 (0.0\%)   & 7 (20.6\%)  & 10 (29.4\%) & 17 (50.0\%) & 10.5 \\
\textit{Performance Optimization}          & 31  & 1 (3.2\%)   & 5 (16.1\%)  & 6 (19.4\%)  & 19 (61.3\%) & 11   \\
\textit{User Interface Design}             & 23  & 7 (30.4\%)  & 3 (13.0\%)  & 8 (34.8\%)  & 5 (21.7\%)  & 5    \\
\textit{Component Dependency Optimization} & 19  & 1 (5.3\%)   & 2 (10.5\%)  & 3 (15.8\%)  & 13 (68.4\%) & 12   \\
\midrule
\textbf{All task instances}                & 376 & 58 (15.4\%) & 106 (28.2\%) & 85 (22.6\%) & 127 (33.8\%) & 5   \\
\bottomrule
\end{tabular}
}
\end{table*}

\textbf{2) What was the purpose of initiating a conversation by the developer?}

Unlike the categories of design tasks explored in RQ1, this aspect, the purpose of initiating a conversation, is to investigate the categories of behaviors and reactions that developers expect from ChatGPT, rather than the categories of their content. For instance, consider an initiating prompt ``\textit{Could you suggest a solution for improving my software system security?}'', RQ1 emphasizes ``system security'', which aligns with the task category of \textit{Security Design}, while this aspect focuses on ``suggesting a solution'', corresponding to the category of \textit{Recommendation of Design Solution} (the results will be presented below). To answer this question, we adopted the Open Coding and Constant Comparison method~\citep{CC2014} to analyze data item D3, which involves the purpose derived from the initial prompt of each task, in order to identify the purposes for which developers initiate conversations. Ultimately, we identified four purposes for developers to initiate a conversation: \textit{Knowledge Query about Design}, \textit{Code Generation for Design}, \textit{Verification of Design Solution}, and \textit{Recommendation of Design Solution}. Below are the details for each purpose, each accompanied by an example:
\begin{itemize}
\item[-] \textbf{\textit{Knowledge Query about Design}}. Inquire about relevant concepts in software design, such as ``\textit{What is the singleton pattern? Please provide an example.}''

\item[-] \textbf{\textit{Code Generation for Design}}. Provide design requirements (such as ``\textit{based on the SOLID principles}'' and ``\textit{ensure \rev{low coupling and high cohesion}}'') and let ChatGPT generate code snippets that meet these design specifications.

\item[-] \textbf{\textit{Verification of Design Solution}}. Present the design solution, code structure, and other relevant details, with the expectation that ChatGPT will identify any flaws and  inconsistencies at the design level. For example, ``\textit{I designed a system to stage files, check it and show me the unreasonable aspects.}''

\item[-] \textbf{\textit{Recommendation of Design Solution}}. From the perspective of requirements, describe the problem and request ChatGPT to generate a design solution. For example, ``\textit{How to implement a platform like IMDb? Show the architecture.}''
\end{itemize}

\indent The distribution of these four purposes of initiating a conversation is presented in Figure~\ref{fig:RQ2Distributions}(a). Since the analysis focuses on the initial prompt of each task, the total count aligns with the number of tasks, amounting to 376. Among the four categories, developers' purposes primarily focus on \textit{Knowledge Query about Design} (121 out of 376, 32.18\%) and \textit{Code Generation for Design} (114 out of 376, 30.32\%). \textit{Recommendation of Design Solution} follows closely (89 out of 376, 23.67\%), while \textit{Verification of Design Solution} is the least emphasized (52 out of 376, 13.83\%).  

\definecolor{rq2purposeknowledge}{HTML}{1F77B4}
\definecolor{rq2purposecode}{HTML}{FF7F0E}
\definecolor{rq2purposerecommend}{HTML}{2CA02C}
\definecolor{rq2purposeverify}{HTML}{D62728}
\definecolor{rq2leveldetailed}{HTML}{9467BD}
\definecolor{rq2levelarchitectural}{HTML}{8C564B}
\definecolor{rq2levelidiom}{HTML}{17BECF}

\begin{figure*}[htbp]
\centering
\begin{minipage}[t]{0.48\linewidth}
\centering
\begin{tikzpicture}[font=\piechartfont\scriptsize, line join=round]
\def\pieRadius{1.85}
\pgfmathsetmacro{\currentAngle}{90}
\foreach \name/\slicevalue/\percentlabel/\slicecolor in {
  {Knowledge Query about Design}/121/32.18/rq2purposeknowledge,
  {Code Generation for Design}/114/30.32/rq2purposecode,
  {Recommendation of Design Solution}/89/23.67/rq2purposerecommend,
  {Verification of Design Solution}/52/13.83/rq2purposeverify%
}{
  \pgfmathsetmacro{\deltaAngle}{\slicevalue/376*360}
  \pgfmathsetmacro{\nextAngle}{\currentAngle+\deltaAngle}
  \pgfmathsetmacro{\midAngle}{\currentAngle+0.5*\deltaAngle}
  \filldraw[fill=\slicecolor, draw=white, line width=0.8pt]
    (0,0) -- (\currentAngle:\pieRadius) arc (\currentAngle:\nextAngle:\pieRadius) -- cycle;
  \ifnum\slicevalue=52
    \node[font=\pieslicelabelfont, inner sep=0pt] at ([xshift=4.8pt,yshift=2.2pt]\midAngle:1.10)
      {\pielabeltext{\slicevalue}{\percentlabel}};
  \else
    \node[font=\pieslicelabelfont] at (\midAngle:1.12)
      {\pielabeltext{\slicevalue}{\percentlabel}};
  \fi
  \xdef\currentAngle{\nextAngle}
}
\end{tikzpicture}

\vspace{0.3em}
\begin{minipage}[c][4.2em][c]{\linewidth}
\centering
{\tikzlegendfont
\begin{tabular}{@{}l@{}}
\textcolor{rq2purposeknowledge}{\rule{1.0ex}{1.0ex}}~Knowledge Query about Design (121) \\
\textcolor{rq2purposecode}{\rule{1.0ex}{1.0ex}}~Code Generation for Design (114) \\
\textcolor{rq2purposerecommend}{\rule{1.0ex}{1.0ex}}~Recommendation of Design Solution (89) \\
\textcolor{rq2purposeverify}{\rule{1.0ex}{1.0ex}}~Verification of Design Solution (52)
\end{tabular}}
\end{minipage}

\vspace{0.25em}
{\scriptsize\normalfont (a) Developers' purposes for initiating a conversation\par}
\end{minipage}\hfill
\begin{minipage}[t]{0.48\linewidth}
\centering
\begin{tikzpicture}[font=\piechartfont\scriptsize, line join=round]
\def\pieRadius{1.85}
\pgfmathsetmacro{\currentAngle}{90}
\foreach \name/\slicevalue/\percentlabel/\slicecolor in {
  {Detailed design level}/203/53.99/rq2leveldetailed,
  {Architectural level}/112/29.79/rq2levelarchitectural,
  {Code idiom level}/61/16.22/rq2levelidiom%
}{
  \pgfmathsetmacro{\deltaAngle}{\slicevalue/376*360}
  \pgfmathsetmacro{\nextAngle}{\currentAngle+\deltaAngle}
  \pgfmathsetmacro{\midAngle}{\currentAngle+0.5*\deltaAngle}
  \filldraw[fill=\slicecolor, draw=white, line width=0.8pt]
    (0,0) -- (\currentAngle:\pieRadius) arc (\currentAngle:\nextAngle:\pieRadius) -- cycle;
  \node[font=\pieslicelabelfont] at (\midAngle:1.12)
    {\pielabeltext{\slicevalue}{\percentlabel}};
  \xdef\currentAngle{\nextAngle}
}
\end{tikzpicture}

\vspace{0.3em}
\begin{minipage}[c][4.2em][c]{\linewidth}
\centering
{\tikzlegendfont
\begin{tabular}{@{}l@{}}
\textcolor{rq2leveldetailed}{\rule{1.0ex}{1.0ex}}~Detailed design level (203) \\
\textcolor{rq2levelarchitectural}{\rule{1.0ex}{1.0ex}}~Architectural level (112) \\
\textcolor{rq2levelidiom}{\rule{1.0ex}{1.0ex}}~Code idiom level (61)
\end{tabular}}
\end{minipage}

\vspace{0.25em}
{\scriptsize\normalfont (b) Software design task levels addressed in the conversation\par}
\end{minipage}
\caption{Distributions of developers' purposes for initiating a conversation and software design task levels addressed in the dialogue}
\label{fig:RQ2Distributions}
\end{figure*}
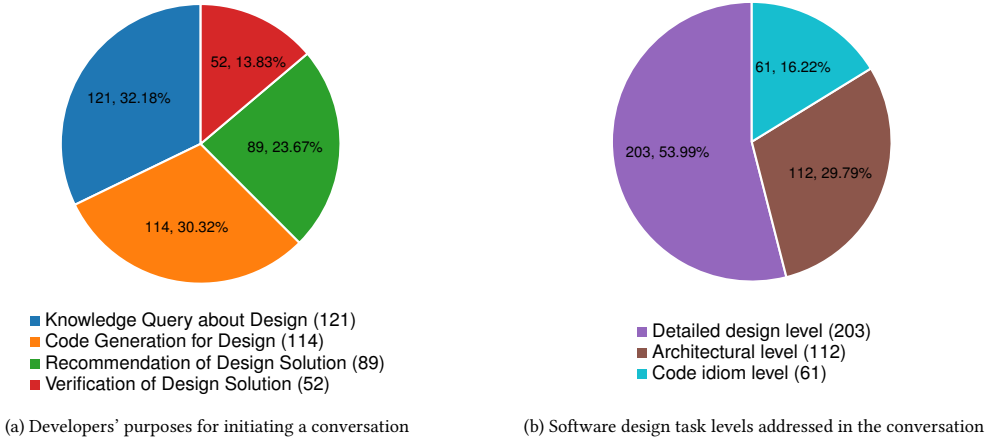

\textbf{3) Which design levels do software design tasks in developer-ChatGPT dialogues most commonly fall into?}

This aspect examines the design levels at which software design tasks in developer-ChatGPT dialogues are concentrated. Unlike the first two aspects,  this aspect focuses on the distribution of task levels across the collected dialogues. Understanding this distribution helps characterize the design levels at which ChatGPT is currently most often used in software design practice. To answer this question, we adopted the Open Coding and Constant Comparison method~\citep{CC2014} to analyze data item D4, which concerns the design level of each software design task discussed in the dialogue. Through iterative coding and comparison, we identified three design levels. We then used the terminology proposed by \citet{PatternOrientedSoftwareArchitecture} to name these levels as follows:
\begin{itemize}
    \item[-] \textbf{\textit{Architectural level}}. Primarily involves high-level, overarching architecture and module information. For example, ``\textit{Can I employ blackboard pattern to coordinate all the data? Alternatively, do you have any suggestions for coordinating them?}''
    \item[-] \textbf{\textit{Detailed design level}}. This mainly refers to the design at the class level, and is manifested in a number of code files involved, with an emphasis on resolving issues in software design. The design patterns proposed by the Gang of Four~\citep{GoF} are at this level. For example, ``\textit{... In this context, which is more advantageous: the command pattern or the chain of responsibility pattern?}''
    \item[-] \textbf{\textit{Code idiom level}}. Code idioms are code fragments for meeting design requirements when implementing a simple task, algorithm, or data structure that is not a built-in feature in the programming language being used. For example, ``\textit{Show me examples to achieve lazy initialization using double-checked locking in Java, and explain the details.}''
\end{itemize}

\indent The distribution of the three levels is presented in Figure~\ref{fig:RQ2Distributions}(b). Among the three levels, the number of tasks at the \textit{Detailed design level} is the highest, with 203 out of 376 (53.99\%), followed by the \textit{Architectural level} with 112 out of 376 (29.79\%), while the \textit{Code idiom level} comprises only 61 tasks (16.22\%).

\subsubsection{Survey Study Results}
Because RQ2 is addressed through the three aspects above, the survey questions SQ5, SQ6, and SQ7 ask participants about the three aspects, respectively. SQ5 is a single-choice survey question, and its total count is equal to the number of participants. SQ6 is a multiple-choice semi-open question, and SQ7 is a multiple-choice \rev{closed-ended} question; consequently, their total counts exceed the number of participants. Responses entered under \rev{``Other''} in SQ6 were manually reviewed and found to fit the predefined categories; they were therefore reassigned to those existing categories from the mining study and included in the total number, with no additional categories emerging from these responses. The responses about the number of dialogue rounds were based on participants' experience.

{\revcolor
The two studies use different units of analysis. The mining study contains 376 task instances, and each instance contributes one prompt purpose and one design level. The survey studdy contains 65 participants; SQ5 has one response per participant, whereas SQ6 and SQ7 allow multiple selections, yielding 120 and 96 selections, respectively. Since these observations are neither independent nor measurements of the same unit, tests of homogeneity and rank correlations are not appropriate. We therefore compare the distributions descriptively. Table~\ref{tab:RQ2ComparisonBetweenMiningAndSurvey} reports percentages within each study. The most prominent prompt purpose and design level are the same in both sources: \textit{Knowledge Query about Design} (32.2\% of task instances; 66.2\% of participants) and \textit{Detailed design level} (54.0\%; 70.8\%). The distributions of dialogue rounds differ: \textit{Over 10 rounds} is the largest mining study group (33.8\%), whereas \textit{4$\sim$10 rounds} is the largest survey group (40.0\%). These are descriptive comparisons and do not support inferential conclusions.
}

\begin{table*}[htbp]
\revcolor
\centering
\caption{\rev{Side-by-side comparison of the RQ2 results from the mining study and the survey study (The unit of analysis differs between the two studies, so the two count columns are not directly comparable; only the relative prominence of the categories within each study is.)}}
\label{tab:RQ2ComparisonBetweenMiningAndSurvey}
\scalebox{0.78}{
\begin{tabular}{llrr}
\toprule
\textbf{Aspect} & \textbf{Category} & \textbf{Mining study} & \textbf{Survey study} \\
 & & \textbf{(task instances,} & \textbf{(participants,} \\
 & & \textbf{$n$~=~376)} & \textbf{$n$~=~65)} \\
\midrule
\multirow{4}{*}{\makecell[l]{Number of dialogue\\rounds (D2 / SQ5)}}
 & Single round                          & 58 (15.4\%)  & 9 (13.8\%)  \\
 & 2$\sim$3 rounds                       & 106 (28.2\%) & 13 (20.0\%) \\
 & 4$\sim$10 rounds                      & 85 (22.6\%)  & 26 (40.0\%) \\
 & Over 10 rounds                        & 127 (33.8\%) & 17 (26.2\%) \\
\midrule
\multirow{4}{*}{\makecell[l]{Purpose of prompt\\(D3 / SQ6)}}
 & \textit{Knowledge Query about Design}     & 121 (32.2\%) & 43 (66.2\%) \\
 & \textit{Code Generation for Design}       & 114 (30.3\%) & 27 (41.5\%) \\
 & \textit{Recommendation of Design Solution}& 89 (23.7\%)  & 31 (47.7\%) \\
 & \textit{Verification of Design Solution}  & 52 (13.8\%)  & 19 (29.2\%) \\
\midrule
\multirow{3}{*}{\makecell[l]{Level of tasks\\(D4 / SQ7)}}
 & \textit{Detailed design level}            & 203 (54.0\%) & 46 (70.8\%) \\
 & \textit{Architectural level}              & 112 (29.8\%) & 31 (47.7\%) \\
 & \textit{Code idiom level}                 & 61 (16.2\%)  & 19 (29.2\%) \\
\bottomrule
\end{tabular}
}

\vspace{0.4em}
{\footnotesize \rev{Percentages in the mining study column are computed over the 376 extracted task instances, each of which contributes exactly one category per aspect. Percentages in the survey study column are computed over the 65 participants. SQ5 is single-choice, so its four counts sum to 65 and its percentages sum to 100\%. SQ6 and SQ7 are multiple-response questions, so a participant may appear in several rows (120 selections for SQ6 and 96 selections for SQ7) and the percentages in those blocks do not sum to 100\%. The row labelled \textit{Over 10 rounds} corresponds to the mining bucket of more than 10 rounds and to the SQ5 answer option worded \textit{More than 10 rounds} in the questionnaire (Table~\ref{tab:survey-questions}).}}
\end{table*}

The \rev{divergence} in the number of dialogue rounds may be explained by the different nature of the two data sources. The mining study counts the observable rounds in concrete task-level ChatGPT conversations, whereas the survey captures participants' retrospective perception of typical LLM use across their own design tasks. Participants may therefore report longer or shorter interaction patterns based on \rev{their overall experience} rather than on a specific completed task. In addition, survey responses aggregate usage across different LLM tools and project contexts, while the mining study is limited to shared ChatGPT conversations on GitHub. We therefore interpret this \rev{divergence} as a difference in perspective between observed task-level behavior and self-reported practice in using LLMs for software design.

{\revcolor
Table~\ref{tab:ExperienceBySQ5} shows that reported dialogue rounds vary by development experience. Twelve of the 29 participants with more than 5 years of experience selected \textit{More than 10 rounds}, compared with one of 12 participants with 1$\sim$3 years of experience and none of the 7 participants with less than 1 year of experience. Participants with less than 3 years of experience most often selected \textit{4$\sim$10 rounds}.

The frequencies of free-text responses also vary across experience groups. All eight mentions of \textit{Contextual reliance} were made by participants with more than 5 years of experience, as were 9 of the 11 mentions of \textit{Limited source file uploading}. Among participants with less than 1 year of experience, \textit{Innovative ideas} was the most frequently reported benefit, accounting for 5 of the 7 mentions. By role, all nine architects selected \textit{Performance Optimization}, whereas none of the 12 testers selected \textit{Architecture Design}, \textit{Security Design}, or \textit{Component Dependency Optimization}. These findings are descriptive and should not be generalized, given the small size of several subgroups.
}

\begin{table*}[htbp]
\revcolor
\centering
\small
\setlength\tabcolsep{6pt}
\renewcommand{\arraystretch}{1.15}
\caption{\rev{Development experience (SQ1) against the reported typical number of dialogue rounds (SQ5) in participants ($n$~=~65) (SQ5 is single-choice, so each participant appears exactly once. Counts are given rather than percentages because two of the four groups contain fewer than 13 participants.)}}
\label{tab:ExperienceBySQ5}
\begin{tabular}{lrrrrr}
\toprule
\textbf{Experience (SQ1)} & \textbf{Single} & \textbf{2$\sim$3} & \textbf{4$\sim$10} & \textbf{$>$10} & \textbf{$n$} \\
\midrule
$<$ 1 year        & 1 & 0 & 6  & 0  & 7  \\
1 $\sim$ 3 years  & 2 & 1 & 8  & 1  & 12 \\
3 $\sim$ 5 years  & 0 & 5 & 8  & 4  & 17 \\
$>$ 5 years       & 6 & 7 & 4  & 12 & 29 \\
\midrule
\textbf{All participants} & 9 & 13 & 26 & 17 & 65 \\
\bottomrule
\end{tabular}
\end{table*}

\begin{tcolorbox}[colback=black!5, colframe=black!20, width=1.0\linewidth, arc=1mm, auto outer arc, boxrule=1.5pt]
{\textbf{RQ2 Summary}: \textit{The analysis of 376 design tasks shows that developers use ChatGPT for software design through both short and long interactions\rev{, with a right-skewed distribution of dialogue rounds (median 5 rounds, IQR 2$\sim$11, mean 6.18, maximum 68, and 33.8\% of the task instances requiring more than 10 rounds). Interaction length differs systematically by task category: \textit{Architecture Design} has a median of 11 rounds and no single-round task instance, whereas \textit{Use of Design Patterns} has a median of 2 rounds and is resolved in a single round in 48.4\% of its task instances}. Developers mainly start conversations to query design knowledge, generate code under design constraints, obtain recommendations of design solutions, or verify existing design solutions. Most tasks are situated at the \textit{Detailed design level}, followed by the \textit{Architectural level} and the \textit{Code idiom level}. \rev{The 65 survey participants report the same ordering of prompt purposes and of design levels, but a shorter typical interaction length than the one observed in the mined conversations.}}}
\end{tcolorbox}

\subsection{RQ3: What are the benefits of using LLMs in software design?}\label{sec:RQ3_results}
We adopted the Open Coding and Constant Comparison method~\citep{CC2014} to analyze the responses from survey question \rev{SQ8, together with the answers to SQ10 that express a benefit}. As a result, we classified the benefits mentioned by the participants when using LLMs for addressing software design issues into 7 categories:

\textbf{(1) Reduced overhead} from design-related tasks, allowing a focus on coding. Participants described LLMs as reducing the burden of early-stage design work and allowing them to focus more on implementation. As \surveyuser{P8} noted, ``\textit{Way more convenient than the typical development process, requirements, documentation, technology selection, and detailed architecture design, or things like these. I can save time from the early stage crappy jobs and concentrate on coding.}'' \surveyuser{P25} similarly reported that LLMs \textit{``free me from miscellaneous tasks and can sometimes produce surprising results.}''

\textbf{(2) Quick project onboarding} for newly joined developers through the clarification of architecture and design. Participants also highlighted the role of LLMs in onboarding and project understanding. For example, \surveyuser{P41} explained that ``\textit{Long context support allows me to input project structures, snippets of source code files, and historical documents into LLMs, making it possible to help me organize the projects. Most of the time, even if LLMs cannot directly produce satisfactory results, they can still help me understand the design rationale behind the project’s history.}'' Likewise, \surveyuser{P32} said that LLMs ``\textit{help me learn about software design without a knowledge barrier}'' and help them understand projects within their company.

\textbf{(3) More effective search engine} when searching for design-related information. Several participants regarded LLMs as an efficient alternative to traditional search engines for design-related knowledge. As \surveyuser{P28} put it, they are ``\textit{Better than search engines when querying knowledge}'', and \surveyuser{P52} likewise described LLMs as ``\textit{nice search engine alternatives}'' that save time.

\textbf{(4) Innovative ideas} for inspiring solutions in software design. Participants also valued the occasional creativity of LLMs in stimulating design ideas. In \surveyuser{P34}'s words, ``\textit{Innovative ideas (though they are rare), well-structured output, human-friendly, just like a great human fellow, and they can notice what I have overlooked}'', while \surveyuser{P11} commented that the responses can be ``\textit{remarkably creative}'', leaving them pleasantly surprised.

\textbf{(5) Early detection} of design flaws and inconsistencies. Another commonly reported benefit was the early detection of flaws and inconsistencies in design solutions. \surveyuser{P23} emphasized that LLMs are ``\textit{very detailed and comprehensive, capable of noticing errors that I haven’t noticed}'', and \surveyuser{P55} noted that uploaded solutions can ``\textit{receive constructive feedback ... until there are no evident logical inconsistencies.}''

\textbf{(6) Technology selection} for overcoming the unfamiliarity of certain technologies and promoting diversity of technology choices in software design. Participants further reported that LLMs support technology selection when teams need to move beyond unfamiliar or outdated stacks. For instance, \surveyuser{P47} explained that their company worked with a very old technology stack and that GPT had ``\textit{assisted many in the selection of a new technology stack.}''

\textbf{(7) Concept interpretation} for reaching a common understanding on the terms used within a team that facilitates collaboration. Finally, some participants reported that LLMs help teams converge on a shared interpretation of design concepts and terminology. \surveyuser{P19} observed that, ``\textit{with complete project documentation}'', LLMs can understand intentions precisely enough to ``\textit{standardize team collaboration}'', and \surveyuser{P3} added that the way LLMs express terminology helps teams reach ``\textit{a relatively unified and definitive interpretation.}''

Figure~\ref{tab:RQ3Statistics} presents the distribution of the seven reported benefits of using LLMs in software design. The most frequently mentioned benefits are \textit{More Effective Search Engine} (25 out of 65 participants, 38.46\%) and \textit{Early Detection} (22 out of 65 participants, \rev{33.85\%}), whereas \textit{Technology Selection} (9 out of 65 participants, 13.85\%) and \textit{Concept Interpretation} (7 out of 65 participants, 10.77\%) were mentioned the least often. 
\rev{Two distinct denominators are used here and should not be confused. The percentages in this paragraph are calculated based on the 65 participants and indicate the proportion of participants who mentioned each benefit. Because participants could mention multiple benefits, a single participant may be counted in multiple categories, causing the percentages to sum to more than 100\%. In contrast, the percentages in Figure~\ref{tab:RQ3Statistics} are calculated based on the 116 benefit mentions coded in total and show how these mentions are distributed across the seven categories.}

\definecolor{benefitoverhead}{HTML}{E76F51}
\definecolor{benefitonboarding}{HTML}{2A9D8F}
\definecolor{benefitsearch}{HTML}{E9C46A}
\definecolor{benefitideas}{HTML}{6D597A}
\definecolor{benefitdetection}{HTML}{4E79A7}
\definecolor{benefittech}{HTML}{B07AA1}
\definecolor{benefitconcept}{HTML}{59A14F}

\begin{figure*}[htbp]
\centering
\begin{tikzpicture}[font=\piechartfont\small, line join=round]
\def\pieRadius{1.85}
\path[use as bounding box] (-6.4,-2.55) rectangle (7.4,2.55);
\pgfmathsetmacro{\currentAngle}{90}
\foreach \name/\slicevalue/\percentlabel/\chartcolor in {
  {More effective search engine}/25/21.55/benefitsearch,
  {Early detection}/22/18.97/benefitdetection,
  {Reduced overhead}/19/16.38/benefitoverhead,
  {Quick project onboarding}/17/14.66/benefitonboarding,
  {Innovative idea}/17/14.66/benefitideas,
  {Technology selection}/9/7.76/benefittech,
  {Concept interpretation}/7/6.03/benefitconcept%
}{
  \pgfmathsetmacro{\deltaAngle}{\slicevalue/116*360}
  \pgfmathsetmacro{\nextAngle}{\currentAngle+\deltaAngle}
  \pgfmathsetmacro{\midAngle}{\currentAngle+0.5*\deltaAngle}
  \filldraw[fill=\chartcolor, draw=white, line width=0.9pt]
    (0,0) -- (\currentAngle:\pieRadius) arc (\currentAngle:\nextAngle:\pieRadius) -- cycle;
  \ifnum\slicevalue<12
    \ifnum\slicevalue=9
      \draw[gray!70, line width=0.35pt] (\midAngle:\pieRadius) -- (1.65,1.84);
      \node[font=\pieslicelabelfont, anchor=west, inner sep=0.6pt] at (1.69,1.84)
        {\pielabeltext{\slicevalue}{\percentlabel}};
    \else
      \draw[gray!70, line width=0.35pt] (\midAngle:\pieRadius) -- (0.62,2.26);
      \node[font=\pieslicelabelfont, anchor=west, inner sep=0.6pt] at (0.66,2.26)
        {\pielabeltext{\slicevalue}{\percentlabel}};
    \fi
  \else
    \node[font=\pieslicelabelfont] at (\midAngle:1.12)
      {\pielabeltext{\slicevalue}{\percentlabel}};
  \fi
  \xdef\currentAngle{\nextAngle}
}
\node[anchor=west, align=left, font=\tikzlegendfont, inner sep=0pt, outer sep=0pt]
  at (2.85,0)
  {
\begin{tabular}{@{}l@{}}
\textcolor{benefitsearch}{\rule{1.0ex}{1.0ex}}~More effective search engine (25) \\
\textcolor{benefitdetection}{\rule{1.0ex}{1.0ex}}~Early detection (22) \\
\textcolor{benefitoverhead}{\rule{1.0ex}{1.0ex}}~Reduced overhead (19) \\
\textcolor{benefitonboarding}{\rule{1.0ex}{1.0ex}}~Quick project onboarding (17) \\
\textcolor{benefitideas}{\rule{1.0ex}{1.0ex}}~Innovative idea (17) \\
\textcolor{benefittech}{\rule{1.0ex}{1.0ex}}~Technology selection (9) \\
\textcolor{benefitconcept}{\rule{1.0ex}{1.0ex}}~Concept interpretation (7)
    \end{tabular}
  };
\end{tikzpicture}
\caption{Distribution of the seven reported benefits of using LLMs in software design \rev{(percentages computed over the 116 coded benefit mentions)}}
\label{tab:RQ3Statistics}
\end{figure*}
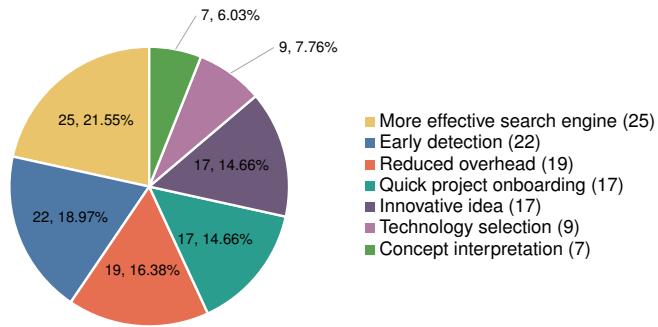

\begin{tcolorbox}[colback=black!5, colframe=black!20, width=1.0\linewidth, arc=1mm, auto outer arc, boxrule=1.5pt]
{\textbf{RQ3 Summary}: \textit{The participants reported seven main benefits of using LLMs for software design. These benefits include reduced overhead from software design tasks, quick project onboarding, and the use of LLMs as a more effective search engine for design knowledge. Participants also reported that LLMs can inspire innovative ideas, support technology selection, help detect design flaws early, and facilitate concept interpretation within a team.}}
\end{tcolorbox}

\subsection{RQ4: What are the limitations of using LLMs in software design?}\label{sec:RQ4_results}
We adopted the Open Coding and Constant Comparison method~\citep{CC2014} to analyze the responses from survey question \rev{SQ9, together with the answers to SQ10 that express a limitation}, and we classified the limitations mentioned by the participants when using LLMs for addressing software design issues into 6 categories:

\textbf{(1) Lengthy outputs} lead to limited readability for design tasks, thereby impacting the practical use of LLMs in software design. One frequently mentioned limitation is the difficulty of reading long LLM responses in design-oriented workflows. As \surveyuser{P13} noted, ``\textit{Chat messages are hard to read}'' and \surveyuser{P27} added that LLMs work well, ``\textit{but for those who use LLMs every day every time, it can be horrible to face such a long chat log.}''

\textbf{(2) Unclear articulation of requirements}  makes LLMs fail \rev{to generate} design solutions that meet the requirements. Participants also pointed out that unclear or imperfectly articulated requirements can easily lead to unsatisfactory design outputs. \surveyuser{P41} said that, ``\textit{Its capacity may be suitable, but communicating with it (or them) in natural language is hard, which is outperformed by human fellows.}'', while \surveyuser{P7} reported that LLMs ``\textit{do not always grasp my meaning accurately, and I need to pay attention to correct their fundamental faults.}''

\textbf{(3) Limited source file uploading} restricts the analysis of projects, resulting in failures to understand the design of the projects. Another limitation concerned the difficulty of providing project files in a form that preserves their structure. \surveyuser{P22} explained that ``\textit{the chat format makes the input of source files very complicated}''. Similarly, \surveyuser{P47} noted that a project with 100 tree-structured files becomes only 20 or 30 ``\textit{flat and disordered ones}'' after uploading.

\textbf{(4) Inexecutable code} generated by LLMs \textbf{from design} hinders \rev{system prototyping and the implementation of design patterns}, typically requiring manual intervention. Participants further observed that LLM-generated code often remains difficult to execute directly in design tasks. \surveyuser{P34} noted that, although LLMs can provide design solutions, they ``\textit{cannot generate even an executable prototype}'', and \surveyuser{P53} added that ``\textit{the code generated still needs manual modifications even if I only asked for a design pattern trick.}''

\textbf{(5) Contextual reliance} makes LLMs unable to handle complex user scenarios and sensitive to missing context in software design. Participants also stressed that LLMs are highly sensitive to missing or ambiguous context. As \surveyuser{P56} warned, ``\textit{If the contextual information is insufficient or the question is phrased somewhat ambiguously, the output of LLMs can become completely unusable.}''

\textbf{(6) Hallucinated results} impact the design quality, making the generated design content irrelevant to requirements. Finally, hallucinated results were reported as directly reducing the practical value of LLM-generated design content. \surveyuser{P62} observed that ``\textit{when there is a lot of content, it is easy to experience hallucinations}'', and \surveyuser{P34} succinctly concluded that ``\textit{hallucinations reduce the practicality of llms.}''

Figure~\ref{tab:RQ4Statistics} presents the distribution of the six reported limitations of using LLMs in software design. The most frequently mentioned limitations were \textit{Lengthy Outputs} (32 out of 65 participants, 49.23\%) and \textit{Inexecutable Code from Design} (28 out of 65 participants, \rev{43.08\%}), whereas \textit{Contextual Reliance} (8 out of 65 participants, \rev{12.31\%}) and \textit{Limited Source File Uploading} (11 out of 65 participants, 16.92\%) were mentioned the least often. 
\rev{As described in Section~\ref{sec:RQ3_results}, the percentages in this paragraph are calculated based on the 65 participants and indicate the proportion who mentioned each limitation. In contrast, the percentages in Figure~\ref{tab:RQ4Statistics} are calculated based on the 118 limitation mentions coded in total and show how these mentions are distributed across the six categories.}

\definecolor{limitlengthy}{HTML}{6BAED6}
\definecolor{limitrequirements}{HTML}{2A9D8F}
\definecolor{limituploading}{HTML}{E9C46A}
\definecolor{limitinexecutable}{HTML}{E76F51}
\definecolor{limitcontext}{HTML}{B07AA1}
\definecolor{limithallucinated}{HTML}{84A59D}

\begin{figure*}[htbp]
\centering
\vspace{-1.5em} 
\begin{tikzpicture}[font=\piechartfont\small, line join=round]
\def\pieRadius{1.85}
\path[use as bounding box] (-2.2,-2.1) rectangle (7.2,2.2);
\pgfmathsetmacro{\currentAngle}{90}
\foreach \name/\slicevalue/\percentlabel/\chartcolor in {
  {Lengthy outputs}/32/27.12/limitlengthy,
  {Inexecutable code from design}/28/23.73/limitinexecutable,
  {Hallucinated results}/21/17.80/limithallucinated,
  {Unclear articulation of requirements}/18/15.25/limitrequirements,
  {Limited source file uploading}/11/9.32/limituploading,
  {Contextual reliance}/8/6.78/limitcontext%
}{
  \pgfmathsetmacro{\deltaAngle}{\slicevalue/118*360}
  \pgfmathsetmacro{\nextAngle}{\currentAngle+\deltaAngle}
  \pgfmathsetmacro{\midAngle}{\currentAngle+0.5*\deltaAngle}
  \filldraw[fill=\chartcolor, draw=white, line width=0.9pt]
    (0,0) -- (\currentAngle:\pieRadius) arc (\currentAngle:\nextAngle:\pieRadius) -- cycle;
  
  \ifnum\slicevalue<12
    \ifnum\slicevalue=11
      \draw[gray!70, line width=0.35pt] (\midAngle:\pieRadius) -- (1.71,1.66);
      \node[font=\pieslicelabelfont, anchor=west, inner sep=0.6pt] at (1.75,1.66)
        {\pielabeltext{\slicevalue}{\percentlabel}};
    \else
      \draw[gray!70, line width=0.35pt] (\midAngle:\pieRadius) -- (0.88,2.08);
      \node[font=\pieslicelabelfont, anchor=west, inner sep=0.6pt] at (0.92,2.08)
        {\pielabeltext{\slicevalue}{\percentlabel}};
    \fi
  \else
    \node[font=\pieslicelabelfont] at (\midAngle:1.12)
      {\pielabeltext{\slicevalue}{\percentlabel}};
  \fi
  \xdef\currentAngle{\nextAngle}
}
\node[anchor=west, align=left, font=\tikzlegendfont, inner sep=0pt, outer sep=0pt]
  at (2.85,0)
  {
\begin{tabular}{@{}l@{}}
\textcolor{limitlengthy}{\rule{1.0ex}{1.0ex}}~Lengthy outputs (32) \\
\textcolor{limitinexecutable}{\rule{1.0ex}{1.0ex}}~Inexecutable code from design (28) \\
\textcolor{limithallucinated}{\rule{1.0ex}{1.0ex}}~Hallucinated results (21) \\
\textcolor{limitrequirements}{\rule{1.0ex}{1.0ex}}~Unclear articulation of requirements (18) \\
\textcolor{limituploading}{\rule{1.0ex}{1.0ex}}~Limited source file uploading (11) \\
\textcolor{limitcontext}{\rule{1.0ex}{1.0ex}}~Contextual reliance (8)
    \end{tabular}
  };
\end{tikzpicture}
\vspace{-1em} 
\caption{Distribution of the six reported limitations of using LLMs in design \rev{(percentages computed over the 118 coded limitation mentions)}}
\label{tab:RQ4Statistics}
\end{figure*}
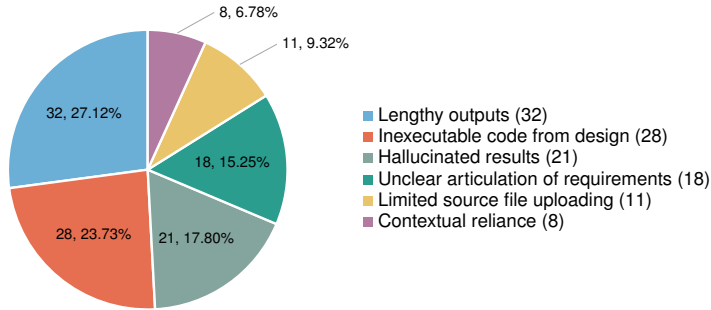

\begin{tcolorbox}[colback=black!5, colframe=black!20, width=1.0\linewidth, arc=1mm, auto outer arc, boxrule=1.5pt]
{\textbf{RQ4 Summary}: \textit{The participants reported six main limitations of using LLMs for software design. These limitations include lengthy outputs, unclear articulation of requirements, and limited source file uploading. Participants also reported inexecutable code, \rev{strong reliance on context}, and hallucinated results, which reduce the quality and usefulness of the generated design outputs.}}
\end{tcolorbox}

\section{Discussions}\label{sec:Discussions & Implications}

\subsection{Interpretation of the Results}
By combining 291 shared ChatGPT conversations with 65 survey responses, our study provides empirical evidence on how developers use ChatGPT and LLMs for software design in practical software development. We interpret the findings by synthesizing the results across RQ1\textasciitilde RQ4.

\subsubsection{ChatGPT is used mainly for design exploration and specific design assistance}
The results of RQ1 and RQ2 show that developers use ChatGPT for a broad range of software design tasks, including \textit{Interface and Protocol Design}, \textit{Architecture Design}, \textit{Data Model Design}, and \textit{Code Refactoring}. However, the dominant purposes are \textit{Knowledge Query about Design} and \textit{Code Generation for Design}, and the most common task level is the \textit{Detailed design level}. Taken together, these findings suggest that the current use of ChatGPT for software design is mainly exploratory and targeted in the sense that developers rely on ChatGPT to explore alternatives, clarify design concepts, and draft candidate solutions for bounded software design problems, rather than to fully \rev{automate} software design decision making. This interpretation is consistent with prior work showing that LLM support for software architecture and design is promising but still fragmented, with outcomes depending strongly on task framing, project context, and evaluation settings~\citep{schmid2025slr,esposito2026genai_sa}. It also helps explain why participants emphasized benefits such as alternative search, quick project onboarding, and early detection of design issues.

\subsubsection{Design-oriented interaction is iterative and strongly context-dependent}
The results of RQ2 indicate that interaction with ChatGPT for software design is rarely a one-shot activity. \rev{A mined task instance requires a median of 5 rounds (IQR 2$\sim$11, mean 6.18), and} both the mining study and the survey study suggest that repeated clarification and refinement are common when developers use ChatGPT or other LLMs for recommendations, code generation, and design exploration. Although the two data sources differ in the exact distribution of dialogue rounds, they converge on a broader pattern: developers often iterate with the model while clarifying requirements and narrowing the solution space. \rev{The breakdown in Table~\ref{tab:RoundsByCategory} further shows that this iteration is not uniform across design tasks but is concentrated where the answer depends on project-specific constraints: \textit{Architecture Design} has a median of 11 rounds and never terminates in a single round, whereas \textit{Use of Design Patterns}, whose answers are named and well documented, has a median of 2.} The limitations reported in RQ4 help explain why this iteration is needed. Unclear articulation of requirements, contextual reliance, and limited support for sharing project artifacts can all reduce the usefulness of the generated output. This observation complements evidence from other SE domains showing that LLM use is often iterative rather than one-shot. Studies on AI-assisted programming show that developers use LLMs for exploration, prompting, inspection, adaptation, and verification of generated outputs~\citep{Barke2023GroundedCopilot,Ross2023ProgrammersAssistant,Liang2024surveyAI,Mozannar2024rbl}. Related work in requirements engineering similarly emphasizes that useful LLM outputs depend on explicit prompts, contextual information, and refinement of requirements-related inputs~\citep{Ronanki2024ReqGenAI,Gorer2023ReqElicitationScripts}. In this sense, prompting becomes part of software design itself because developers must externalize assumptions~\citep{yang2018assumptions}, constraints~\citep{tang2009modeling}, and rationale~\citep{falessi2013value} in a form that the model can process. The need for iteration and context refinement also varies depending on the type of design task at hand. For example, architectural decisions often demand a broader understanding of project scope, dependencies, and long-term maintenance considerations~\citep{tofan2013difficulty,li2022understanding}. In contrast, detailed design tasks may require more precise contextual information to ensure that generated solutions are technically feasible and adhere to project-specific constraints~\citep{diazpace2024archmind}. Thus, the iterative nature of LLM interaction adapts to the varying degrees of complexity and the level of abstraction inherent in diverse software design tasks.

\subsubsection{Perceived efficiency gains are bounded by verification and integration costs}
The results of RQ3 show that \rev{the survey participants perceive} substantial practical value in using LLMs for software design, especially for reducing overhead, accelerating project familiarization, and obtaining rapid access to design knowledge. At the same time, the results of RQ4 highlight frictions that can offset these gains, such as lengthy outputs, hallucinated results, and code that still requires manual correction before it becomes usable. Together, these findings suggest that LLMs currently create the most value in accelerating the exploration of the design space and helping developers better understand design requirements, constraints, and generated design suggestions, whereas the responsibility for checking feasibility, consistency, and implementation fitness remains with developers. This interpretation is compatible with recent evidence from AI-assisted programming showing that net productivity gains may diminish or disappear when verification and integration overhead become substantial in complex engineering tasks~\citep{metr2025rct}. For software design, where requirements are often underspecified and output quality is difficult to assess immediately, such overheads may be especially important.

\rev{We explicitly acknowledge the limitations of this comparison. The data behind RQ3 and RQ4 are free-text reports of perceived benefits and limitations. They are not designed to measure time, effort, or rework, and they contain no counterfactual against which a productivity effect could be estimated. Our study can therefore neither corroborate nor contradict the result of the randomized trial reported by~\cite{metr2025rct}, and we cite that work solely to motivate treating verification and integration overhead as serious concerns, not as evidence that these costs offset the gains reported by our participants. Determining whether they do would require directly measuring design outcomes and developer effort, which we identify as future work in Section~\ref{sec:Conclusions}.}

\subsubsection{\rev{The design tasks brought to ChatGPT differ from those emphasized in the research literature}}
{\revcolor
Our results differ from the main emphasis of the software design literature reviewed in Section~\ref{sec:LLMsForSDIssues}. Prior studies commonly evaluate requirement-to-design translation, UML or class diagram generation, and the generation or recovery of architectural knowledge~\citep{acm_uml_eval_2024,nomad_uml_2025,cot_classdiagram_2024,Dhar2024ADD,diazpace2024archmind,zhou2025using}. In our corpus, \textit{Interface and Protocol Design} (19.4\%) and \textit{Use of Design Patterns} (16.5\%) are the largest categories. \textit{Detailed design level} accounts for 54.0\% of task instances, whereas \textit{Architecture Design} accounts for 13.8\%. None of the nine categories directly represent requirement-to-design translation as evaluated in those studies.

Two factors may explain this difference. Interface and pattern related design tasks can often be described and assessed with limited project context, whereas architecture tasks depend on system wide information. Participants reported \textit{Limited source file uploading} and \textit{Contextual reliance}, which may restrict the architecture tasks they can address through a chat interface. In addition, the conversations were shared in GitHub code, commits, issues, pull requests, and discussions, where design work associated with an implementation change is more likely to appear than earlier architecture decisions.

These findings suggest that evaluations should include design tasks observed in practice and provide realistic project context. Benchmarks focused only on isolated requirements-to-diagram prompts may not represent the distribution of tasks in our corpus.
}

\subsection{Implications}
{\revcolor
We organize the implications around three roles: researchers, who design studies, benchmarks, and evaluations of LLMs for software design; practitioners, who use LLMs in their software design work; and tool builders, who develop the interfaces and integrations through which practitioners interact with LLMs. These roles may overlap and should not be interpreted as mutually exclusive or as representing separately sampled populations. We discuss the implications for each role in turn, drawing on the findings of this study.
}

\textbf{For researchers}, our findings suggest that future work on LLMs for software design should evaluate tasks that reflect actual practice, such as \textit{Interface and Protocol Design}, \textit{Architecture Design}, \textit{Data Model Design}, and \textit{Code Refactoring}. Such evaluations should go beyond whether a generated artifact looks plausible. They should also examine requirements coverage, consistency of design decisions, downstream rework, and the effort required from developers to verify and adapt model outputs. In addition, more in-situ and longitudinal studies are needed to understand how factors such as prompt specificity, available project context, model choice, and iteration depth influence the quality and usefulness of LLM-supported software design work~\citep{schmid2025slr,esposito2026genai_sa}.
\rev{Such longitudinal designs could also assess how the use of LLMs in software design evolves over time and impacts long-term software quality.} It is essential to explore how iterative interactions with LLMs affect both the design process and the quality of generated designs. Such studies could reveal whether the initial benefits observed in rapid prototyping and design refinement translate into lasting improvements or lead to unforeseen challenges as projects evolve.

\textbf{For practitioners}, the results suggest that LLMs are currently most useful as exploratory support for software design rather than as substitutes for design expertise. They appear well suited for clarifying concepts, comparing candidate solutions, drafting initial artifacts, and obtaining early feedback on tentative designs. However, developers should treat generated outputs as provisional and should subject them to domain review, architectural reasoning, and implementation checks before refinement and adoption. Our findings on contextual reliance and articulation of requirements also suggest that \rev{better performance on design related tasks may be more likely when prompts include explicit constraints}, relevant project background, and clear evaluation criteria~\citep{white2023promptpatterns}.
While LLMs provide valuable assistance in design exploration and prototyping, it is critical for developers to retain control over final design decisions. Human expertise remains essential in interpreting the generated design solutions and ensuring they align with project-specific requirements and constraints. Practitioners need to strike a careful balance between automating design space exploration and the generation of candidate solutions, and applying human judgment to verification, integration, and final decision-making. Consequently, LLM-generated designs \rev{should} undergo rigorous review before being incorporated into production systems.

\textbf{For tool builders}, our results point to opportunities to reduce the gap between conversational interaction and actionable support for software design. Participants' comments on limited source-file uploading and lengthy outputs suggest that design-oriented tools should better support structured project context, cross-file navigation, and output formats that are easier to inspect, compare, and revise. Rather than presenting only long chat responses, future tools could help developers transform LLM outputs into software-design artifacts such as API sketches, Architecture Decision Record drafts, schema proposals, or reviewable refactoring plans. Support for linking LLM suggestions to project artifacts and preserving traceability across iterations may also reduce the burden of verifying LLM-generated design outputs identified in RQ4.

\section{Threats to Validity}\label{sec:Threats}
Drawing on common validity considerations in empirical software engineering~\citep{Wohlin2012ESE}, we discuss the threats to the validity of this study that are related to construct validity, internal validity, external validity, and reliability.

\subsection{Construct Validity}
Construct validity refers to whether the study captures the phenomenon of interest in an appropriate way. A first issue relates to how we operationalize ``software design'' when collecting and labeling the shared ChatGPT conversations. Some borderline cases between software design, coding, and general development problem solving are inherently ambiguous. To mitigate this issue, we defined explicit inclusion and exclusion criteria for the mining study, conducted the pilot extraction, and resolved uncertain cases through discussion among multiple authors.


A second threat arises from the use of proxy measures. For example, the number of dialogue rounds helps characterize interaction patterns, but it does not directly measure task complexity, task success, or solution quality. Similarly, the benefits and limitations reported in RQ3 and RQ4 reflect practitioners' perceptions rather than objective productivity or long-term project outcomes. We therefore interpret these results as perceived usage characteristics of LLMs for software design rather than as causal evidence of effectiveness.

A third threat involves the survey instrument. Several survey items for RQ1 and RQ2 were derived from categories identified in the mining study, which may have anchored participants toward our categorization. We reduced this risk by using semi-open questions for SQ4 and SQ6, allowing participants to add missing options, and by using open-ended questions for RQ3 and RQ4. 

{\revcolor
SQ4, SQ6, and SQ7 present categories derived from the mining study as answer options. Participants could therefore confirm the categories and introduce a different category through an ``Other'' response. Seven participants used ``Other'' when answering SQ4, and we subsequently assigned those responses to existing categories. We accordingly treat the survey as complementary evidence rather than independent validation.

The two studies may also measure different constructs under the same label. The survey options were presented without the operational definitions used in the mining study, so participants may have interpreted terms such as \textit{Architecture Design} or \textit{Detailed design level} differently. The questionnaire and its Welcome page are included in the replication package~\citep{replpack}. Future studies may provide definitions and examples for each option or elicit categories before presenting a fixed list.
}

\subsection{Internal Validity}
Internal validity relates to the credibility of the interpretations drawn from the data and the extent to which alternative explanations may account for the observed results. Our study is observational and cross-sectional, so it does not support causal claims about the effects of using LLMs on software design quality or developer productivity. Reported benefits and limitations may be influenced by factors such as developer experience, task complexity, project domain, or organizational context, and the survey results reflect practitioners' experience and perceptions rather than directly measured outcomes.

Another challenge to internal validity is the evolving nature of the studied technology. The mining dataset spans from May 2023 to January 2025, and the survey was conducted in June 2025. During this period, ChatGPT capabilities, interfaces, and usage practices evolved considerably. As a result, some observed interaction patterns may reflect temporal variation in the tool rather than stable properties of LLM-supported software design work. This temporal difference may also partly explain why the mining study and the survey study were not fully consistent regarding the numbers of dialogue rounds. To avoid over-interpretation, we used the two studies in a complementary way and treated their relationship as triangulation rather than as direct confirmation of the same underlying pattern.

{\revcolor
The mining data were collected over 20 months during which ChatGPT models and interfaces changed. A conversation shared in mid-2023 may therefore reflect more limited context and file-handling capabilities than a conversation shared in late 2024 or January 2025. Round counts and reports of limited source-file uploading may depend on these changes. We treat each conversation as an interaction with the version of ChatGPT available at that time and interpret the aggregate as a summary of a heterogeneous period. We do not assume that the distributions are stable across model generations or attribute the difference between mining and survey round counts to a single cause.
}

\subsection{External Validity}
External validity concerns the generalizability of our findings. The mining study relies on publicly shared ChatGPT conversation links on GitHub, and the survey participants were recruited through GitHub and LinkedIn. These sources provide access to active developer communities, but they also form a convenience sample and may not represent developers working in private repositories, regulated domains, or organizations with different policies for using LLMs.

{\revcolor
Both data sources are self-selected. The mining dataset includes conversations that developers chose to share publicly in GitHub artifacts; private, unsuccessful, or routine interactions may therefore be underrepresented. The survey received 65 valid responses from 1,130 invitations (5.8\%). Respondents may have had stronger experience or opinions about LLM use than non-respondents. We report the sample composition by recruitment channel, experience, and role, but the results should be interpreted as observations from these samples rather than prevalence estimates for a wider practitioner population.
}

In addition, our results focus on ChatGPT and on conversations collected during a specific period. Therefore, the findings may not transfer directly to other LLMs, other development tools, or later generations of LLM-based assistants for software design.

Coverage of the mining dataset also introduces a limitation. Our GitHub-based retrieval was constrained by platform search behavior, the API limit of 1,000 items per query, and the availability of shared links at the time of collection. Therefore, the dataset should not be interpreted as exhaustive. We frame the mining study as an empirical characterization of observed use of ChatGPT for software design rather than as a census of all such interactions.

\subsection{Reliability}
Reliability concerns whether the study procedures are sufficiently transparent and stable to support replication. Qualitative coding and manual classification inevitably involve researcher judgment. We mitigated this threat by documenting the data collection and analysis procedure, conducting the pilot extraction, involving multiple authors in key classification and coding steps, and resolving disagreements through discussion. In addition, we have made the replication package~\citep{replpack} publicly available, including the mining study dataset, the survey materials, and the survey responses, to facilitate the replication of our study.

{\revcolor
 After the double-coded pilot, the first author performed the extraction and initial coding of the full dataset, with uncertain cases discussed by the first three authors. We therefore cannot report an agreement coefficient for the 291 conversations or 376 task instances. Independent double coding of the full dataset could provide additional evidence about category stability. We make the assignment process inspectable by reporting operational definitions and boundary rules and by releasing each task instance with its conversation content, extracted data, and assigned codes in the replication package~\citep{replpack}.
} 

\section{Conclusions and Future Work}\label{sec:Conclusions}
This paper presents a mixed-methods study of how developers use ChatGPT and other LLMs for software design, combining a mining study of 291 shared ChatGPT conversations with a survey of 65 practitioners. Across the two data sources, we found that developers use LLMs for a diverse set of design tasks, and that these interactions are usually multi-turn and are \rev{initiated} mainly for \textit{Knowledge Query about Design}, \textit{Code Generation for Design}, and \textit{Recommendation of Design Solution}. The evidence also shows that the current use of LLMs for software design is concentrated at the \textit{Detailed design level}, although architectural tasks are also common.

The benefits reported by practitioners indicate that LLMs can reduce overhead from design-related tasks, support quick project onboarding, provide rapid access to design knowledge, and help uncover potential design issues. However, the reported limitations show that practical usage of LLMs for software design remains constrained by lengthy outputs, incomplete project context, hallucinated results, and inexecutable code generated by LLMs, which hinders \rev{system prototyping and the implementation of design patterns}. Taken together, these results suggest that LLMs are currently more useful for design exploration and iterative design refinement than for fully automated software design.

{\revcolor
 The conclusions concern two self-selected samples: 291 conversations shared publicly in GitHub artifacts and 65 practitioners who responded to 1,130 invitations. They characterize observed and reported practice in these samples. They do not estimate the prevalence of design tasks in the wider practitioner population, establish that the nine categories are exhaustive, or show that the categories are the only valid organization of the data. The consensus-coding procedure and its limitations are discussed in Section~\ref{sec:Threats}.
}

Overall, our study provides empirical insight into how developers use LLMs for software design in practice, including the design tasks involved, the ways developers interact with ChatGPT, and the key benefits and limitations of such use. The results of this study suggest several promising directions for future research. First, replications across other LLMs and industrial settings are needed to assess the robustness of the design task categories, developer-ChatGPT interaction patterns, benefits, and \rev{limitations observed across RQ1\textasciitilde RQ4}. Second, longitudinal or controlled studies can be employed to investigate how the usage of LLMs affects design quality and labor cost, \rev{and to evaluate the quality of LLM-generated design outputs and the effort required to verify them.} Third, future tools should better integrate project artifacts, preserve repository structure, and output in formats that are easier to inspect, compare, and revise, such as drafts of architecture decision records, design diagrams, or reviewable architecture refactoring plans.

\section*{Data availability}
The replication package of this study, including the mining study dataset, the survey instrument, and the survey responses, has been made available at~\citep{replpack}.

\section*{Declaration of competing interest}
\rev{The authors declare that they have no known competing financial interests or personal relationships that could have appeared to influence the work reported in this paper.}

\section*{Declaration of generative AI and AI-assisted technologies in the writing process}
\rev{During the preparation of this work, the authors used generative AI tools solely to improve the language and readability of the manuscript. After using these tools, the authors reviewed and edited the content as needed and take full responsibility for the content of the publication. Generative AI was not used to produce, analyze, or interpret the study data.}

\section*{Acknowledgments}
This work has been partially supported by the National Natural Science Foundation of China (NSFC) with Grant No. 92582203 and 62402348.

\printcredits

\bibliographystyle{cas-model2-names}
\bibliography{ref}
\balance

\end{sloppypar}
\end{document}